%Paper: hep-th/9305057
%From: "MICHAEL P. TUITE" <MPHTUITE@bodkin.ucg.ie>
%Date: Thu, 13 May 1993 12:38 GMT

%%%%%%%%%%%%%%%%%%%%%%%%%%%%%%%%%%%%%%%%%%%%%%%%%%%%%%%%%%%%%%%%%
%								%
%    On the Relationship between the Uniqueness of the 	        %
% 	Moonshine Module and Monstrous Moonshine 		%
%								%
%		Michael P. Tuite				%
%								%
%   Requires Harvmac TeX macros available from HEPTH Network	%
%								%
%%%%%%%%%%%%%%%%%%%%%%%%%%%%%%%%%%%%%%%%%%%%%%%%%%%%%%%%%%%%%%%%%

\input harvmac
\magnification=1200
\hsize=15.5truecm
\overfullrule=0pt
\def\beginsection #1 {\vskip 1.5 truecm
\line{\bf #1 \hfil}\smallskip\message {#1}}

\def \sqr#1#2{\vcenter
{\hrule height .#2pt \hbox{\vrule width .#2pt
height #1pt\kern#1pt\vrule width .#2pt}\hrule height .#2pt}}
\def\minimatrix#1{\null\,\vcenter{\normalbaselines\mathsurround=0pt
    \ialign{\hfil$##$\hfil&&\enspace\hfil$##$\hfil\crcr
      \mathstrut\crcr\noalign{\kern-\baselineskip}
      #1\crcr\mathstrut\crcr\noalign{\kern-\baselineskip}}}\,}

\def \textorbsqr#1#2{\hbox{$\minimatrix{{#1}&\sqr{8}{8}\cr
                                  {}&{#2}\cr}$}}

\def \orbsqr#1#2{\hbox{$\minimatrix{{ }&{ }\cr
                                {#1}&\sqr{12}{12}\cr
                                {}&{#2}\cr}$}}

\def \MMorbsqr#1#2{\hbox{$\minimatrix{{ }&{ }\cr
                                {#1}&\sqr{12}{12}^{\ \natural}\cr
                                {}&{#2}\cr}$}}
\def \MMtextorbsqr#1#2{\hbox{$\minimatrix{{ }&{ }\cr
                                {#1}&\sqr{8}{8}^{\ \natural}\cr
                                {}&{#2}\cr}$}}

\def \subsect#1{\medskip\noindent {\bf #1}}

\def\pfu{partition function\ }

\def\aut{automorphism}
\def\auts{automorphisms}
\def\auto{automorphism\ }
\def\autos{automorphisms\ }
\def\mod{modular\ }
\def\Th{Thompson series\ }

\def\orb{orbifold\ }
\def\op{operator\ }
\def\ops{operators\ }

\def\Tr#1#2{{\rm Tr}_{#1}(#2)}
\def\Torb#1{T_{#1}^{\rm orb}(\tau)}
\def\det#1{{\rm det}(#1)}

\def\MM{{\cal V}^{\natural}}
\def\a{{\overline a}}
\def\b{{\overline b}}
\def\c{{\overline c}}
\def\g{{\overline g}}
\def\r {{\overline r}}
\def\Tgt{T_g(\tau)}
\def\L{\Lambda}
\def\HMM{{\cal H}^\natural}
\def\HL{{\cal H}_\L}
\def\Hr{{\cal H}_r}
\def\Ha {{\cal H}_a}
\def\Hah {{\cal H}_{a^h}}
\def\Hb {{\cal H}_b}
\def\Hg {{\cal H}^\natural_g}

\def\VL{{\cal V}^\L}
\def\VLtilde{\tilde{\cal V}^\L}
\def\VLa{\tilde{\cal V}^\L_a}

\def\VLb{\tilde{\cal V}^\L_b}
\def\VLc{{\cal V}_c}
\def\VLrc{{\cal V}_{rc}}
\def\Vr{{\cal V}_r}
\def\Va{{\cal V}_a}
\def\Vb{{\cal V}_b}
\def\Vc{{\cal V}^\natural_c}
\def\Vp {{\cal V}'}
\def\Vak{{\cal V}_{a^k}}
\def\Vg{{\cal V}^\natural_g}
\def\Vgk{{\cal V}^\natural_{g^k}}

\def\Hgh{{\cal H}^\natural_{g^h}}
\def\Vah {{\cal V}_{a^h}}
\def\Vorb{{\cal V}_{\rm orb}^a}
\def\Vorbh{{\cal V}_{\rm orb}^{a'}}
\def\Vp{{\cal V}'}
\def \Pap{{\cal P}_{a'}}
\def \Vap{{\cal V}_{a'}}
\def \Vbp{{\cal V}_{b'}}
\def \Vf{{\cal V}_{f}^\natural}
\def \Vfr{{\cal V}_{f^r}^\natural}
\def \Vfk{{\cal V}_{f^k}^\natural}
\def \Vfone{{\cal V}_{f_1}^\natural}
\def \Vftwo{{\cal V}_{f_2}^\natural}

\def \Vfr{{\cal V}_{f^r}^\natural}
\def \Vfe{{\cal V}_{f^e}^\natural}
\def \Vfhat{{\cal V}_{\hat f}^\natural}
\def\MMorb{{\cal V}_{\rm orb}^g}
\def\Morb{M_{\rm orb}^a}
\def\Morbh{ M_{\rm orb}^{a^h}}
\def\Horb{{\cal H}_{\rm orb}^a}

\def \MMforb{{\cal V}_{\rm orb}^{f}}

\def \MMfrorb{{\cal V}_{\rm orb}^{f^r}}
\def\para{{\vert\vert}}
\def\Lr{L_\r}
\def\La{L_\a}
\def\Lb{L_\b}
\def\Lrhat{\hat\Lr}
\def\Lbhat{\hat\Lb}
\def\Lahat{\hat\La}
\def\Pr{{\cal P}_r}
\def\Pa{{\cal P}_a}

\def\Pg{{\cal P}_g}
\def \Tft {T_f(\tau)}
\def \Pf {{\cal P}_{f}}
\def \Pfr {{\cal P}_{f^r}}
\def\Co{{\rm Co}_1}
\def\G{\Gamma}
\def\Gg{\G_g}
\def\Ga{\G_a}
\def\GN{\G_0(N)}
\def\Gn{\G_0(n)}
\def\Gp{\G_0(p)}
\def\Gnh{\G_0(n\vert h)}

\def\Gnhe{\Gnh+e_1,e_2,...}
\def\Nor{{\cal N}(\GN)}
\lref\FLM{
Frenkel, I.,  Lepowsky, J.  and Meurman, A.,
{\it A moonshine module for the monster} in:
J.Lepowsky et al. (eds.), Vertex operators in mathematics and physics,
(Springer Verlag, New York, 1985).}

\lref\FLMb{
Frenkel, I.,  Lepowsky, J.  and Meurman, A.,
Vertex operator algebras and the monster,
(Academic Press, New York, 1988).}

\lref\FLMzero{
Frenkel, I.,  Lepowsky, J.  and Meurman, A.,
{\it {A natural representation of the Fischer-Griess monster
with the modular function J as character}},
\break Proc.Natl.Acad.Sci.USA {\bf 81} (1984) 3256.}

\lref\Dixon{
Dixon, L.,   Harvey, J.A., Vafa, C.,  and Witten, E.
{\it Strings on orbifolds},
Nucl.Phys.\break {\bf B261} (1985) 678;
{\it Strings on orbifolds II},
Nucl.Phys. {\bf B274} (1986) 285.}

\lref\DGH{
Dixon, L. Ginsparg, P.  and Harvey, J.A.,
{\it Beauty and the beast: superconformal symmetry in a monster module},
Comm.Math.Phys. {\bf 119} (1988) 285.}

\lref\CS{
Conway, J.H.  and Sloane,  N.J.A.,
Sphere packings, lattices and groups,
(Springer Verlag, New York, 1988).}

\lref\GO{
Goddard, P.  and Olive, D.,
{\it Algebras, lattices and strings} in:
J.Lepowsky et al. (eds.), Vertex operators in mathematics and physics,
(Springer Verlag, New York, 1985).}

\lref\GSW{
Green, M.   Schwarz, J., and  Witten, E.,
Superstrings Vol 1,
(Cambridge University Press, Cambridge, 1987).}

\lref\Serre{
Serre, J-P.,
A course in arithmetic,
(Springer Verlag, New York, 1970).}

\lref\Goddard{
Goddard, P.,
{\it Meromorphic conformal field theory} in:
Proceedings of the CIRM Luminy conference, 1988,
(World Scientific, Singapore, 1989).}

\lref\DVVV{
Dijkgraaf, R.,   Vafa, C., Verlinde,  E. and  Verlinde, H.,
{\it The operator algebra of orbifold models},
Comm.Math.Phys. {\bf 123} (1989) 485.}

\lref\CHZtwo{
Corrigan, E.   and  Hollowood, T.J.,
{\it Comments on the algebra of straight,
 twisted and intertwining vertex operators},
Nucl.Phys. {\bf B304} (1988) 77.}

\lref\DGMZtwo{
Dolan, L.,  Goddard, P. and Montague, P.,
{\it Conformal field theory of twisted vertex operators},
Nucl.Phys. {\bf B338} (1990) 529.}

\lref\DFMS{
Dixon, L.,  Friedan,  D., Martinec, E.  and Shenker,  S.,
{\it The conformal field theory of orbifolds},
Nucl.Phys. {\bf B282} (1987) 13.}

\lref\Vafa{
Vafa, C.,
{\it Modular invariance and discrete torsion on orbifolds},
Nucl.Phys.\break {\bf B273} (1986) 592.}

\lref\FV{
Freed, D. and Vafa,  C.,
{\it Global anomalies on orbifolds},
Commun.Math.Phys. {\bf 110} (1987) 349.;
{\it Comment on "Global anomalies on orbifolds"},
Commun.Math.Phys. {\bf 117} (1988) 349.}

\lref\Ginsparg{
Ginsparg, P.,
{\it Applied conformal field theory}, Les Houches,
Session XLIX, 1988, "Fields, strings and critical
phenomena", ed. E. Brezin and J. Zinn-Justin,
 Elsevier Science Publishers (1989).}

\lref\BPZ{
Belavin, A.A.,  Polyakov,  A.M. and A.B. Zamolodchikov, A.B.,
{\it Infinite conformal symmetry and two dimensional quantum field theory},
Nucl.Phys. {\bf B241} (1984) 333.}

\lref\KacP{
Kac, V.  and Peterson, D.,
{\it 112 constructions of the basic representation of the loop group of $E_8$},
Proceedings of the Argonne symposium
on anomolies, geometry, topology, 1985 (World Scientific, Singapore, 1985).}

\lref\Griess{
Griess, R.,
{\it The friendly giant},
Inv.Math. {\bf 68} (1982) 1.}

\lref\Thompson{
Thompson, J.G.,
{\it Finite groups and modular functions},
Bull.London Math.Soc.\break{\bf 11} (1979) 347.}

\lref\Tuitetwo{
Tuite, M.P.,
{\it 37 new orbifold  constructions  of the Moonshine module ?},
DIAS-STP-90-30  To appear in Commun.Math.Phys.}

\lref\Tuitegen{
Tuite, M.P.,
To appear.}

\lref\DongMason{
Dong, C.  and Mason, G.,
{\it On the construction of the mooonshine module as a $Z_p$ orbifold},
U.C.Santa Cruz Preprint 1992.}

\lref\CN{
Conway, J.H. and  Norton, S.P.,
{\it Monstrous Moonshine},
Bull.London.Math.Soc.\break {\bf 11} (1979) 308.}

\lref\Atlas{
Conway, J.H., Curtis, R.T.,  Norton, S.P.,  Parker, R.A. and  Wilson, R.A.,
An atlas of finite groups,
(Clarendon Press, Oxford, 1985).}

\lref\Lepowsky{
Lepowsky, J.,
{\it Calculus of twisted vertex operators},
Proc.Natl.Acad.Sci.USA {\bf 82} \break(1985) 8295.}

\lref\CHZn{
Corrigan, E.   and  Hollowood, T.J.,
{\it A  bosonic representation  of the twisted string emission vertex},
Nucl.Phys. {\bf B303} (1988) 135.}

\lref\Kondo{
Kondo, T.,
{\it The automorphism group of the Leech lattice and
elliptic modular functions},
J.Math.Soc.Japan {\bf 37} (1985) 337.}

\lref\Borch{
Borcherds, R.,
{\it Monstrous moonshine and monstrous Lie superalgebras},
Univ. Cambridge DPMMS preprint 1989.}

\lref\Wilson{
Wilson, R.,
{\it The maximal subgroups of Conway's group ${\rm Co}_1$},
J.Alg. {\bf 85} (1983) 144.}

\lref\Narain{
Narain, K.S.,
{\it New heterotic string theories in uncompactified dimensions $< 10$},
Phys. Lett.  {\bf 169B} (1986) 41.}

\lref\Mason{
Mason, G. (with an appendix by Norton, S. P. ),
{\it Finite groups and modular functions},
Proc.Symp.Pure Math. {\bf 47} (1987) 181.}

\lref\DGMtri{
Dolan, L.,  Goddard, P.  and Montague, P.,
{\it Conformal field theory, triality and the monster group},
Phys.Lett. {\bf B236} (1990) 165.}

\lref\Tuiteone{
Tuite, M.P.,
{\it Monstrous  moonshine from orbifolds},
Commun.Math.Phys. {\bf 146} (1992) 277.}

\lref\Ogg{
Ogg, A.,
{\it Hyperelliptic modular curves},
Bull.Soc.Math.France {\bf 102} (1974) 449.}

\lref\Gunning{
Gunning, R.C.,
Lectures on modular forms,
(Princeton University Press, Princeton, 1962).}

\lref\Mont{
Montague, P.S.,
{\it Discussion of self-dual c=24 conformal field theories},
DAMTP preprint,  June 1992.}

\lref\Lang{
Lang, M-L.,
{\it On a question raised by Conway and Norton},
J.Math.Soc.Japan {\bf 41} (1989) 265.}

\lref\KondoTasaka{
Kondo, T.  and Tasaka, T.,
{\it The theta functions of sublattices of the Leech lattice},
Nagoya Math.J. {\bf 101} (1986) 151;
{\it The theta functions of sublattices of the Leech lattice II},
J.Fac.Sci.Univ.Tokyo {\bf 34} (1987) 545.}

\lref \GOKM{
Goddard, P.  and Olive, D.,
{\it Kac-Moody  and Virasoro algebras in relation to quantum physics },
Int.J.Mod.Phys. {\bf A1} (1986) 303.}

\lref\Queen{
Queen, L.,
{\it Modular functions arising from some finite groups},
Math.Comp. {\bf 37} (1981) 547.}

\lref\Norton{
Norton, S.P.,
{\it Generalised moonshine},
 Talk presented at Moonshine workshop, Glasgow, 1992.}

\hfill DIAS-STP-93-09
\smallskip
\hfill May 1993
\vskip 1truecm
\centerline {\bf ON THE RELATIONSHIP BETWEEN MONSTROUS MOONSHINE}
\centerline {\bf  AND THE UNIQUENESS OF THE MOONSHINE MODULE}
\vskip 2truecm
\centerline{Michael P. Tuite \footnote *{EMAIL: mphtuite@bodkin.ucg.ie}}
\centerline{Department of Mathematical Physics}
\centerline {University College,}
\centerline {Galway, Ireland}
\smallskip
\centerline {and}
\smallskip
\centerline{Dublin Institute for Advanced Studies}
\centerline{10 Burlington Road}
\centerline{Dublin 4, Ireland}
\vskip 2truecm
\centerline{\bf ABSTRACT}

\noindent We consider the relationship between the conjectured
uniqueness of the Moonshine Module, ${\cal V}^\natural$,
and Monstrous Moonshine, the genus zero property of the modular
invariance group for each Monster group Thompson series.
We first discuss a family of possible  $Z_n$ meromorphic orbifold
constructions of ${\cal V}^\natural$ based on
\autos of the Leech lattice compactified bosonic string.
We reproduce the Thompson series for all 51 non-Fricke classes of the
Monster group $M$ together with a new
relationship between the centralisers of these classes and 51
corresponding Conway group centralisers (generalising a well-known relationship
 for 5 such classes).  Assuming that ${\cal V}^\natural$
is unique, we then consider meromorphic orbifoldings of ${\cal V}^\natural$ and
show that Monstrous Moonshine holds
 if and only if the only meromorphic orbifoldings of  ${\cal V}^\natural$
are  ${\cal V}^\natural$ itself or the Leech theory.
This constraint on the meromorphic orbifoldings of ${\cal V}^\natural$
 therefore relates Monstrous Moonshine to the uniqueness of
${\cal V}^\natural$ in a new way.

\eject
\centerline{$\phantom{\pi}$}
\vskip 2truecm
\pageno=1

\beginsection {1. Introduction.}
The Moonshine Module, $\MM$, of Frenkel, Lepowsky and Meurman
 (FLM) \refs{\FLMzero,\FLM,\FLMb} is historically
the first example of a $Z_2$ orbifold model
\Dixon\ in Conformal Field Theory (CFT) \refs{\BPZ,\Ginsparg}.
The orbifold construction  is based on  a reflection \auto
of the central charge 24 bosonic string which has been compactified
\Narain\ via the  Leech lattice  cf.\CS.
The vertex \ops (primary conformal fields) of $\MM$ form a closed meromorphic
Operator Product Algebra (OPA) \refs{\FLMb,\Goddard,\DGMZtwo}
which is preserved by the Fischer-Griess Monster group, $M$ \Griess.
 By construction,
$\MM$ has no massless (conformal dimension 1) \ops and has modular
invariant partition function $J(\tau)$, the unique modular invariant
meromorphic function with a simple
pole at $\tau=\infty$ and no constant term.
$J(\tau)$ is unique because the fundamental region for the
full modular group is of genus zero cf.\Serre. Conway and
Norton \CN\ conjectured that this genus zero property
extends to other modular functions called the Thompson series
$\Tgt$ for each conjugacy class of $g\in M$ \Thompson.
Such  a genus zero modular function is called a hauptmodul
and this conjecture that each $\Tgt$  is a hauptmodul is
referred to as Monstrous Moonshine.  Borcherds \Borch\  has now proved
the Moonshine conjectures but the origin of the genus zero property
is still unclear.
One of the main purposes of this paper is to provide a
derivation of  Monstrous Moonshine from
a new principle related to the FLM uniqueness conjecture for $\MM$ which
states that $\MM$ is the unique central charge 24 meromorphic
CFT (up to isomorphisms) with partition function $J(\tau)$ \FLMb.
 Recently, Dong and Mason \DongMason\
have provided rigorous $Z_p$ meromorphic orbifold  constructions
based on prime order $p$  \autos of  the Leech theory for $p=3,5,7,13$
each  with partition function $J(\tau)$.
The resulting CFTs have been proved to be isomorphic to
$\MM$ for $p=3$ and almost certainly so for $p=5,7,13$,
 lending weight to the FLM uniqueness conjecture.

This work is broadly divided into two parts. In the first part (\S2 and \S3)
we discuss further evidence for the uniqueness of $\MM$
where a family of $Z_n$ meromorphic orbifoldings of the
 Leech theory (including the 5 prime
ordered ones) which possibly reproduce $\MM$  are described \Tuitetwo.
We also argue that each such candidate construction of $\MM$
can be reorbifolded to reproduce the Leech theory again.
 In the second part of the
paper, in \S4, we discuss other meromorphic orbifoldings of
$\MM$ with respect to $g\in M$ \Tuiteone. We show that given the
uniqueness of $\MM$, then this orbifolding of $\MM$ can give only $\MM$ or the
Leech theory if and only if the corresponding Thompson series is of genus zero.
Thus, assuming the uniqueness of $\MM$, Monstrous Moonshine can
be derived from the constraints on the possible meromorphic orbifoldings
of $\MM$. The advantage of our approach is that a natural interpretation
for a Thompson series is given and the origin of
the modular invariance group for each series is clearly understood.
Furthermore, when we show that Monstrous Moonshine
is equivalent to the above constraints
on the meromorphic orbifoldings of $\MM$ (given the uniqueness of $\MM$),
 a case by case study of the classes of $M$ is not required.

We begin in \S 2 with a review of both the FLM construction of
$\MM$ \refs{\FLMzero,\FLM,\FLMb} from the point of view of CFT
\refs{\BPZ,\Ginsparg,\DFMS,\DVVV,\DGH}\  and Monstrous Moonshine \CN.
In \S 3 a family of $Z_n$ meromorphic
orbifoldings of the Leech theory (including the 5 prime
ordered ones) based on 38 \autos of the Leech lattice
are described each with partition function $J(\tau)$ so that each orbifolding
is a  candidate construction of $\MM$ \Tuitetwo.
Extensive use of non-meromorphic OPAs for various twisted \op sectors is made
in both \S 2 and \S 3 since such algebras provide the most natural
setting for describing
orbifold constructions \refs{\DFMS,\DVVV}.  However, it must be
stated that a fully rigorous description of non-meromorphic
 OPAs  has yet to be provided.
We show that for each $Z_n$ meromorphic orbifolding of the Leech
theory there is
a corresponding  reorbifolding with respect to a \lq dual \aut\rq\
 which reproduces the Leech theory again so that
the Leech theory is an orbifold partner to each such construction.
Within these constructions, we naturally reproduce
$\Tgt$  of genus zero
 for all of the 51  non-Fricke elements of $M$ i.e. $\Tgt$
is not invariant under the Fricke involution $\tau\rightarrow -1/nh\tau$, $h$
an integer. We also find a
generalisation of an observation of Conway and Norton \CN\  (for prime order
 $p=2,3,5,7,13$) relating the centralisers of the non-Fricke  elements in $M$
to corresponding centralisers in the Conway group, the \auto group of the
Leech lattice.  Finally, we explicitly find for 11 of the
38 orbifold constructions,  a
$Z_2$ reorbifolding which reproduces the Leech theory again and hence,
as recently argued by Montague \Mont, these constructions
must be equivalent to
$\MM$. All of these results strongly indicate that each $Z_n$
 construction reproduces $\MM$ and that $\MM$ is indeed unique and that
the Leech theory is the orbifold partner
to $\MM$ for the meromorphic orbifoldings
of $\MM$ with respect to non-Fricke elements
in $M$. In \S4 we consider meromorphic orbifoldings of $\MM$
 with respect to the remaining
Fricke elements of $M$. We show that assuming
the FLM uniqueness conjecture for $\MM$,
then a meromorphic orbifolding of  $\MM$ with respect
to an element $g\in M$  reproduces $\MM$  (i.e. $\MM$ is an orbifold partner
to itself) if and only if
$\Tgt$  is of genus zero and is Fricke invariant. This result relies on the
analysis of \Tuiteone\ where we related Monstrous Moonshine to the
vacuum properties of $g\in M$ twisted  operators.
A standard construction of these twisted sectors is explicitly
described for elements related to Leech lattice \autos but
otherwise, we assume such twisted \op sectors exist. We also
 assume in all cases
that these \ops satisfy a closed  non-meromorphic OPA.
Together with the results of \S 3, we therefore find that,
assuming $\MM$ is unique,
then $\MM$ has either only itself or the Leech theory as a meromorphic
 orbifold partner if
and only if Monstrous Moonshine holds for Thompson series. In Appendix A
we review the modular groups required to describe Monstrous Moonshine. In
Appendix B we discuss a subgroup of the \auto group for the
 OPA of a $Z_n$ orbifolding of the Leech theory.  This group is required to
express the centraliser relationship between $M$ and the Conway group
described in \S 3.

\beginsection {2. The Moonshine Module and Monstrous Moonshine}

\subsect {2.1 Introduction. }
In this section we review the construction of the Moonshine Module,
denoted by $\MM$,
of Frenkel, Lepowsky and Meurman (FLM) \refs{\FLMzero,\FLM,\FLMb} in the
language of conformal field theory (CFT) \refs{\BPZ,\Ginsparg,\DGH}.
We emphasise certain aspects of this construction which we will later refer to
both in considering possible alternative constructions of $\MM$ in \S 3 and
\lq reorbifoldings\rq\ of $\MM$ in \S 4.
We also review the main feature of this theory which is that the \auto
group of $\MM$ is the Monster group $M$, the largest sporadic finite simple
group. Finally, we introduce the Thompson series
\Thompson\ for $g\in M$ which is the object of interest
in the work of Conway and Norton known as \lq Monstrous Moonshine\rq\ \CN.

The Moonshine Module is a $Z_2$ orbifold CFT \Dixon\ and is
based on a Euclidean closed bosonic
string compactified to a 24 dimensional torus $T^{24}$ \Narain.
The torus $T^{24}$ chosen is that defined by quotienting $R^{24}$ with
the Leech lattice which we denote throughout by $\Lambda$.
$\Lambda$ is the unique 24 dimensional even self-dual Euclidean lattice without
roots i.e. $\langle \alpha,\alpha\rangle \not = 2 $ cf. \refs{\CS,\GO}.
The $Z_2$ orbifolding construction is then based on the
 reflection \auto of $\Lambda$.

 \subsect{2.2 The Leech lattice string construction. }
We begin with the usual left-moving closed bosonic string variables $X^i(z)$
where  $z=\exp(2\pi (\sigma_0+i\sigma_1))$ parameterises the string world sheet
with \lq space\rq\ coordinate $0\le\sigma_1\le 1$ and \lq time\rq\ coordinate
$\sigma_0$ \GSW. On the torus $T^{24}$  the closed string boundary
condition is $X^i(e^{2\pi i}z)=X^i(z)+2\pi\beta^i$ for $\beta\in\Lambda$.
   The standard mode expansion for $X^i(z)$ is
$$
\eqalign{X^i(z)=q^i-ip^i{\rm ln}z+i\sum_{m\neq 0}{\alpha^i_m\over m}z^{-m}}
\eqno(2.1)
$$
with commutation relations
$$
\eqalign{[q^i,p^j]&=i\delta^{ij}\cr
[\alpha^i_m,\alpha^j_n]&=m\delta^{ij}\delta_{m+n,0}}
\eqno(2.2)
$$
A similar expansion holds for  the right-moving part
of the string $X^i(\overline z)$.
The 1-loop \pfu corresponding to a world sheet torus
$z\sim e^{2\pi i}z \sim e^{2\pi i\tau}z$  is
parameterised by the modular parameter $\tau$ with ${\rm Im\ }\tau > 0$.
Since $\Lambda$ is even self-dual, the \pfu  factorizes into
$Z(\tau)Z(\overline\tau)$ where $Z(\tau)$ is a modular invariant function
$$
\eqalign{Z(\tau)={\rm Tr}(q^{L_0})=
{\Theta_\Lambda(\tau)\over {\eta^{24}(\tau)}}}
\eqno{(2.3)}$$
with $q=e^{2\pi i\tau}$ and
where $\Theta_\Lambda(\tau)=\sum_{\beta\in\Lambda}q^{\beta^2/2}$
is the theta function  associated with the Leech lattice $\Lambda$
and is a modular form of weight 12 \Serre.
$L_0={1\over 2}p^2+\sum_{m=1}^\infty \alpha_{-m}^i\alpha_m^i-1$
is the normal ordered Virasoro Hamiltonian operator and
$\eta(\tau)=q^{1\over24}\prod_n(1-q^n)$ is the Dedekind eta function
arising from the oscillator modes. The normal ordering
constant gives the usual bosonic tachyonic
vacuum energy $-1$ for central charge $24$.

The set of primary conformal fields  or vertex operators for this
theory also factorizes into
meromorphic in $z$ (anti-meromorphic in $\bar z$) pieces which form a local
meromorphic  (anti-meromorphic) \op product algebra (OPA).
We will consider the left-moving string which forms a meromorphic
CFT \Goddard. The associated set of primary conformal fields, denoted by
$\VL$, consists of normal ordered vertex \ops $\{\phi(z)\}$ of the form
$$
\eqalign{
\phi_{n_1...n_r}^{i_1...i_r}(\beta,z)=
:\del_z^{n_1}X^{i_1}(z)...\del_z^{n_r}X^{i_r}(z)
e^{i\langle \beta,X(z)\rangle }:c(\beta)}
\eqno{(2.4)}
$$
with integer conformal dimension $h_{\phi}=n_1+...+n_r+\beta^2/2$
where $c(\beta)$ is the standard \lq cocycle factor\rq\
necessary for a local meromorphic OPA \refs{\FLMb,\GO,\DGMZtwo}
$$
\eqalign{
\phi_i(z)\phi_j(w)=\phi_j(w)\phi_i(z)
&\sim \sum_k C^{\phi \phi}_{ijk}(z-w)^{h_k-h_i-h_j}\phi_k(w)+...}
\eqno(2.5)
$$
The first equality in (2.5),
which is the locality condition, relies on a suitable analytic continuation
from $\vert z\vert>\vert w\vert$ to $\vert z\vert<\vert w\vert$.
The cocycle factors in (2.4) are elements of a section of a central extension
$\hat\Lambda$ of $\Lambda$ by $\pm 1$ and obey
$$
\eqalignno{
c(\alpha)c(\beta) c^{-1}(\alpha)c^{-1}(\beta)=
&(-1)^{\langle \alpha,\beta\rangle }&(2.6a)\cr
c(\alpha)c(\beta)=&\epsilon(\alpha,\beta)c(\alpha+\beta)&(2.6b)\cr
\epsilon(\alpha,\beta)\epsilon(\alpha+\beta,\gamma)=&
\epsilon(\alpha,\beta+\gamma)\epsilon(\beta,\gamma)&(2.6c)\cr}
$$
The commutator (2.6a) defines the central extension whereas
$\epsilon(\alpha,\beta)\in\{\pm 1\}$
of (2.6b) is a two-cocycle which depends on the section of $\hat\Lambda$ chosen
and must obey the cocycle condition (2.6c).
Let us denote the Hilbert space of states associated with $\VL$,
$\{\vert\phi\rangle =\lim_{z\rightarrow 0}\phi(z)\vert 0\rangle \}$, by $\HL$.
These states can equivalently be constructed as a Fock
space by the action of creation
\ops $\{\alpha^i_{-n}\}$, $n>0$, on the highest weight states given by
$\{\vert \beta\rangle\} $ where $p^i\vert \beta\rangle =
\beta^i\vert \beta\rangle $.
The trace in (2.3) is then performed over $\HL$.
$Z(\tau)$ is a meromorphic and modular invariant function of $\tau$ with a
unique simple pole at $q=0$ due to the tachyonic vacuum energy. $Z(\tau)$ is
therefore given by the unique (up to an additive constant)
modular invariant function $J(\tau)$ as follows
$$
\eqalignno{Z(\tau)&=J(\tau)+24&(2.7a)\cr
J(\tau)&={E_2^3(\tau)\over \eta^{24}(\tau)}-744
={1\over q}+0+196884q+...&(2.7b)\cr}
$$
where $E_2(\tau)$ is the Eisenstein modular form of weight 4 \Serre. Since $\L$
contains no roots, there are only 24 massless
(conformal dimension 1) \ops $\del_z X^i(z)$.

The FLM Moonshine Module  \refs{\FLMzero,\FLM,\FLMb}
is an \orb CFT \refs{\Dixon,\DFMS}
based on the $Z_2$ lattice reflection
\auto $\r:\beta\rightarrow -\beta$ for $\beta\in\Lambda$.
The elements of $\VL$ form a projective representation of the \auto
group of $\Lambda$, the Conway group ${\rm Co}_0$, due to the cocycle factors
of (2.6) \refs{\FLMzero,\FLMb}. Thus the \auto group of $\VL$ which
preserves the OPA (2.5) is a central extension
$2^{24}.{\rm Co}_0$ of ${\rm Co}_0$ by $Z_2^{24}$  (where $2^{24}$ denotes
$Z_2^{24}$ and where $A.B$ denotes
a group with normal subgroup $A$ and quotient group $B=A.B/A$).  In particular,
the lattice \auto  $\r$ lifts to a set of $2^{24}$ \autos of $\VL$.
With the cocycle factors chosen so that
 $\epsilon(\alpha,\beta)=\epsilon(-\alpha,-\beta)$
 we can define a distinguished lifting of $\r$ to $r$  by
$$
\eqalignno{
rc(\beta)r^{-1}=&c(-\beta) &(2.8a)\cr
r\del_z X^i(z)r^{-1}=& -\del_z X^i(z)&(2.8b)\cr
}
$$
which respects (2.5) and (2.6). Defining the projection
 \op $\Pr=(1+r)/2$, we let $\phi^{(+)}(z)=\Pr\phi(z)$
and $\phi^{(-)}=(1-\Pr)\phi(z)$ be
$\pm 1$ eigenvectors of $r$. The set of \ops $\{\phi^{(+)}\}=\Pr \VL$ then
also form a meromorphic OPA. However, the corresponding \pfu
$\Tr{\Pr\HL}{q^{L_0}}=
{1\over 2}(\textorbsqr{1}{1}+\textorbsqr{r}{1})$ is not modular invariant,
 employing the standard notation for the world-sheet torus boundary conditions
e.g. \refs{\Ginsparg, \DVVV}.
Thus, under a \mod transformation $S:\tau\rightarrow -1/\tau$,
$$
\eqalign{\orbsqr{r}{1}={1\over\eta_\r(\tau)}\rightarrow
\orbsqr{1}{r}=2^{12}[{\eta(\tau)\over\eta(\tau/2)}]^{24}}
\eqno(2.9)
$$
where $\eta_\r(\tau)=[\eta(2\tau)/\eta(\tau)]^{24}$. Therefore a
\lq twisted\rq\ sector $\Hr$ is introduced
to form a modular invariant theory \refs{\FLMzero,\Dixon,\DFMS}.

\subsect{2.3 The twisted string construction.}
Consider a closed string field $\tilde X^i(z)$
obeying the $\r$ twisted boundary condition (monodromy condition)
$\tilde X(e^{2\pi i}z)=-\tilde X(z)+2\pi\beta$, $\beta\in \Lambda$
with mode expansion
$$
\eqalign{
\tilde X^i(z)=
\tilde q^i+i\sum_{m\in Z+{1\over 2}}{\tilde\alpha^i_m\over m}z^{-m}
}
\eqno(2.10)
$$
where the oscillator modes obey the same commutator relations as given in (2.1)
and $\tilde q^i \in \Lr=\Lambda/2\Lambda$,  the $\r$ fixed point space of the
torus. Then $\Lr=Z_2^{24}$ which we
denote by $2^{24}$. The states
$\{\vert \psi\rangle \}$ of the twisted sector $\Hr$ can
again be constructed from a set
of vertex \ops $\VLtilde$ acting, in this case, on a degenerate twisted vacuum.
$\Hr$ can be also constructed as a Fock space
from the action of creation \ops $\{\tilde\alpha^i_{-m}\}$, $m>0$, on
this  degenerate vacuum. These states are
graded by the twisted Virasoro Hamiltonian
$ L_0=\sum_{m\in Z+1/2}^\infty\tilde\alpha_{-m}^i\tilde\alpha_m^i+{1\over2}$
with half integer energies
where the normal ordering constant is now $1\over 2$.
The resulting \pfu  is then $\Tr{{\cal H}_r} {q^{L_0}}=\textorbsqr{1}{r}$
of (2.9).

For each $\phi(z)\in \VL$ there is a corresponding
\op $\tilde\phi(z)\in\VLtilde$, with the same conformal dimension, which is
physically interpreted as the emission of an untwisted state
from the twisted vacuum.
$\VLtilde$  then provides a representation of the OPA for $\VL$ which
 is non-meromorphic because of half integer grading
\refs{\FLMb,\DFMS,\DGMZtwo}.
The construction of $\tilde\phi(z)$ is similar to (2.4) where the
cocycle factors are replaced by a finite set of matrices, $\{c_T(\beta)\}$,
acting on the degenerate twisted vacuum with $\beta$
 a representative element of $\Lr$ where
$\beta\sim\alpha\iff\beta-\alpha\in 2\Lambda$.
These cocycle matrices are defined as follows. The commutator map (2.6a) also
defines a central extension $\Lrhat$ of $\Lr$ by
$\pm 1$.  Then $\Lrhat = 2^{1+24}_{+}$, which denotes an extra-special group
of the given order (with the defining property that the centre $\{\pm 1\}$
and commutator subgroup coincide). There exists a unique faithful irreducible
$2^{12}$  dimensional representation $\pi$
of $\Lrhat$  in which the centre of $\Lrhat$ is represented by $\pm 1$
\refs{\FLMb,\KacP}. The elements of $\pi(\Lrhat)$ are
the twisted cocycle matrices $\{c_T(\beta)\}$ and the vacuum states,
$\{\vert \sigma^l_r\rangle \}$, $l=1,...2^{12}$, form a basis for
the vector space
on which $\pi(\Lrhat)$ acts. These cocycle factors are again necessary for the
twisted vertex \op modes to possess well-defined commutation relations
\refs{\FLMb,\GOKM,\CHZtwo,\DGMZtwo}.
$\pi(\Lrhat)$ can be constructed from appropriate Dirac matrices since
the elements of $\Lrhat$ form a Clifford algebra \refs{\GO,\DGMZtwo}.

The defining characteristic of the \ops $\{\tilde\phi(z)\}$ which act on the
degenerate vacuum states $\{\vert \sigma^l_r\rangle \}$ is the monodromy
condition associated with $r$
$$
\eqalign{\tilde\phi^{(\pm)}(e^{2\pi i}z)=
r^{-1}\tilde\phi^{(\pm)}(z)r}
\eqno(2.11)
$$
where $r\tilde\phi^{(\pm)}(z)r^{-1}= \pm\tilde\phi^{(\pm)}(z)$
as defined above on the corresponding untwisted \op $\phi^{(\pm)}(z)$
e.g. $\del_z\tilde X^i(e^{2\pi i}z)=-\del_z\tilde X^i(z)$.
Using the principles of CFT \BPZ,
each vacuum state $\vert \sigma^l_r\rangle $ is created from the
untwisted vacuum by a primary \lq twist\rq\ conformal field (intertwining \op)
$\sigma^l_r(z)$ with conformal dimension $3\over 2$ where
$\vert \sigma^l_r\rangle =\lim_{z\rightarrow 0}\sigma^l_r(z)\vert 0\rangle $
 \refs{\FLMb, \CHZtwo,\DGMZtwo}.
{}From (2.11), these \ops form a non-meromorphic OPA with the vertex \ops of
$\VL$ and $\VLtilde$ \refs{\DFMS,\CHZtwo,\DGMZtwo} as follows
$$
\eqalign{
\tilde\phi^{(\pm)}(z)\sigma^l_r(w)=\sigma^l_r(w)\phi^{(\pm)}(z)  \sim
(z-w)^{h_\psi-h_{\phi}-3/2}\psi^{(\mp)}(w)+...
}
\eqno{(2.12)}
$$
with a suitable analytic continuation assumed in the first equality.
$\psi^{(\mp)}(w)$ is a primary conformal \op which creates a higher conformal
dimension $h_\psi$ twisted state from the untwisted vacuum where (2.11) implies
that $h_{\psi^{(+)}}\in Z,\ h_{\psi^{(-)}}\in Z+1/2$
e.g. the first excited twisted states
$\vert \psi^{(+)}_{il}\rangle = \tilde\alpha^i_{-1/2}\vert\sigma^l_r\rangle $
with $h_{\psi}=2$ are given by
$\lim_{z\rightarrow 0}z^{1/2}\del_z\tilde X^i(z)\vert \sigma^l_r\rangle $,
the action of the first excited \ops of $\VLtilde$.
We denote the set of \ops $\{\psi(z)\}$, which includes $\{\sigma^l_r(z)\}$,
 by $\Vr$.

The lattice \auto $\r$ also lifts to a set of \autos of $\Vr$.
Since $\Lr$ is invariant under $\r$, $\r$ is
lifted to $\pm 1$ in its action on the degenerate vacuum. We choose the
lifting,
which we also denote by $r$, to be
$$
\eqalignno{
r\sigma^l_r(z)r^{-1}=& -\sigma^l_r(z)&(2.13a)\cr
r\psi^{(\pm)}(z)r^{-1}=&
\pm \psi^{(\pm)}(z)=e^{-2\pi i h_\psi}\psi^{(\pm)}(z)&(2.13b)\cr}
$$
which preserves the OPA (2.12) so that the \ops
with integer valued conformal weights are invariant
under $r$. Then (2.11) and (2.12) imply that the twisted \ops $\Vr$
 when acting on the vacuum $\vert 0\rangle $ obey the monodromy condition
$$
\eqalign{\psi(e^{2\pi i}z)=e^{-2\pi i h_\psi}\psi(z)=r \psi(z)r^{-1}}
\eqno{(2.14)}
$$
(2.14) implies that under the modular transformation $T:\tau\rightarrow\tau+1$,
 $\textorbsqr{1}{r}\rightarrow \textorbsqr{r^{-1}}{r}$.
Thus the lifting of $\r$ chosen in
(2.13) is compatible with the twisted vacuum energy of $1/2$  and ensures
that no extra phase occurs in this transformation.

The OPA (2.12) can be generalised by replacing  $\sigma^l_r(w)$ by any
twisted state $\psi(w)\in\Vr$. Likewise, we may define for each
$\psi(z)\in\Vr$ a vertex \op $\tilde\psi(z)\in\tilde\Vr$ which acts on the
twisted vacuum to give an untwisted  state. The set of such operators forms
a closed non-meromorphic OPA \refs{\FLMb,\DFMS,\CHZtwo,\DGMZtwo}
$$
\eqalignno{
\tilde\phi_i(z) \psi_j(w)
&\sim  \sum_{k} C^{\phi \psi}_{ijk}(z-w)^{h_k-h_i-h_j}\psi_k(w)+...
& (2.15a)\cr
\tilde\psi_i(z) \psi_j(w)
&\sim  \sum_{k} C^{\psi \psi}_{ijk}(z-w)^{h_k-h_{i}-h_{j}}\phi_k(w)+...
 & (2.15b)\cr}
$$
$\VL$ is thus enlarged by the inclusion of the twist fields $\{\sigma^l_r(z)\}$
to $\Vp=\VL\oplus\Vr$ which forms a closed non-meromorphic OPA. Furthermore,
the $r$ invariant set $\Pr\Vp$ forms a closed meromorphic OPA and defines
a modular invariant meromorphic CFT.  This
 is the FLM Moonshine Module  $\MM$ \refs{\FLMzero,\FLM, \FLMb}.
As far as we are aware, a completely rigorous
construction of (2.15) does not yet exist except for this $\Pr$
projection which forms a  meromorphic OPA.
This projection ensures the
absence of the 24 massless (conformal dimension 1) \ops
$\partial_zX^i(z)$ whereas the twisted sector \ops are all massive
since the twisted vacuum energy is $1/2$. Therefore,
the modular invariant \pfu for the associated Hilbert space of states $\HMM$
is
$$
\eqalign{
\Tr{\HMM}{q^{L_0}}=\orbsqr{\Pr}{1}+\orbsqr{\Pr}{r}=J(\tau)}
\eqno (2.16)
$$
where $J(\tau)$ is the unique modular invariant of (2.7b)
without a constant term.

The absence of any
massless \ops in $\MM$ is the crucial feature that sets the Moonshine Module
apart from any other string theory. Normally such \ops
are present and form a Kac-Moody algebra. However, in the present case,
the 196884 conformal dimension 2 operators, including the energy-momentum
tensor $T(z)=-{1\over 2}:\partial_z X^i(z)\partial_z X^i(z):$, can be used to
 define a closed non-associative
commutative algebra. FLM  \refs{\FLMzero,\FLM,\FLMb}  showed that
this algebra is  an affine version of the 196883 dimensional Griess algebra
 \Griess\ together with the energy-momentum tensor.
The \auto group of the Griess algebra
is the Monster finite simple group $M$ of order
 $2^{46}.3^{20}.5^9.7^6.11^2.13^3.17.19.23.29.31.41.47.59.71\sim
8\times10^{53}$.
FLM further showed that $M$ is the \auto group for the full OPA
of $\MM$ where $T(z)$ is a singlet. Thus the \ops of $\MM$ of
a given conformal weight form (reducible) representations of $M$.
This explains
an earlier observation of McKay and Thompson \Thompson\ that the coefficients
of the modular function $J(\tau)$ are
positive sums of dimensions of the irreducible
representations of $M$ e.g. the coefficient of
 $q$ is $196884=1+196883$, the sum
of the trivial and adjoint representation formed by the Griess algebra.

\subsect{2.4 A Monster group centraliser and $\bf Z_2$
reorbifolding $\bf \MM$.}
We may identify an involution (order two) \auto $i\in M$,
defined like a \lq fermion number\rq,
under which all untwisted (twisted) \ops have eigenvalue $+1( -1)$. $i$
clearly also respects the larger non-local OPA of
(2.5) and (2.15). The centraliser
$C(i\vert M)=\{ g\in M\vert ig=gi\}$
may also be determined since this is given by
all OPA \autos which map $\Pr\VL$ and $\Pr\Vr$ into themselves.
As stated earlier,
the \auto group of $\VL$ consists of all liftings of the Conway group
${\rm Co}_0$  to \autos of the OPA (2.5) and is given by
$\Lr.{\rm Co_0}$ where $\Lr=2^{24}$. The fixed point space  $\Lr$ is
invariant under $\r$ and
the \auto group of the twisted sector $\Vr$ is $\Lrhat.\Co$,
an extension of the Conway simple group $\Co={\rm Co}_0/\{1,\r\}$
by $\Lrhat=2^{1+24}_{+}$. The extension is determined by the \auto group of the
twisted cocycle matrices $c_T(\alpha)\in \pi(\Lrhat)$.
In particular, the inner \autos of $\pi(\Lrhat)$ defined by
$c_T(\alpha):c_T(\beta)\rightarrow
c_T(\alpha)c_T(\beta)c_T(\alpha)^{-1}=
(-1)^{\alpha.\beta}c_T(\beta)$ describe  the liftings of the identity element
of
$\Co$ and the given extension. The \auto group for $\Vr$ then
follows from (2.12). Putting these \auto groups together, one can show that the
corresponding \auto group for the projected set of \ops $\Pr\Vp$ is
$C(i\vert M)=2^{1+24}_{+}.\Co$ (see Appendix B).
This result is an essential part of the FLM
construction since Griess showed that $M$ is generated by
$2^{1+24}_{+}.\Co$ and a second involution  $\sigma$. FLM constructed
$\sigma$, which mixes the untwisted and twisted sectors, from a hidden triality
OPA symmetry in the theory \refs{\FLMzero,\FLM,\FLMb,\DGMtri}
and so demonstrated that the \auto group
of $\MM$ is $M$.

The \autos $i$ and $r$ can be said to be
\lq dual\rq\ to each other in the sense that  both are \autos of
 the non-meromorphic OPA for
$\Vp=\VL\oplus\Vr$ and that the subsets invariant under $i$ and $r$, $\VL$ and
$\MM$ repectively, form  meromorphic OPAs. Then
 we may \lq reorbifold\rq\ $\MM$ with respect to $i$ by employing the
24 massless \ops $\{\del_zX^i(z)\}$  to re-introduce the $r=-1$ eigenvalue \ops
$\{\phi^{(-)}\}\oplus\{\psi^{(-)}\}$ where (schematically)
$\phi^{(+)}\del_zX\sim\phi^{(-)},\ \psi^{(+)}\del_zX \sim\psi^{(-)}$ from (2.5)
and (2.15) i.e. the \ops $\{\del_zX^i(z)\}$ create the states
of the $i$ twisted
vacuum. Similarly, monodromy conditions analogous to (2.11) and (2.14) also
hold with $r$ replaced by $i$, $\VL$ replaced by $\MM$ in (2.11) and
$\Vr$ replaced by $\{\phi^{(-)}\}\oplus\{\psi^{(-)}\}$ in (2.12).
{}From this point of view
the two meromorphic constructions $\VL$ and $\MM$ are placed on an
equal footing with each contained in the enlarged set $\Vp$ and each related
to the other by an appropriate $Z_2$ orbifolding procedure.  Equivalently,
we can define $\Vp$ to be the set of all \ops which form a meromorphic OPA with
 $\Pr\VL={\cal P}_i\MM$ i.e. $\Vp$ is \lq dual\rq\ to $\Pr\VL={\cal P}_i\MM$
in the sense suggested by Goddard \Goddard.
The orbifolding of $\VL$ with respect to $r$ is
then $\MM=\Pr\Vp$ and the orbifolding of $\MM$ with respect to $i$ is
$\VL={\cal P}_i\Vp$ i.e.
$$
\eqalign{
\matrix{
 \Vp  \cr
{\buildrel {\cal P}_{i}\ \  \over\swarrow} \qquad
 {\buildrel \ \ \Pr\over\searrow}  \cr
 \VL\quad
         \matrix{{\buildrel r\over \longrightarrow}  \cr
                {\buildrel i\over \longleftarrow}}
\quad\MM\cr}
}
\eqno (2.17)
$$
where the horizontal(diagonal) arrows denote an orbifolding(projection).

\subsect{2.5 Thompson Series, Hauptmoduls and Monstrous Moonshine.}
The states of $\HMM$ of a given conformal weight
form reducible representations of the Monster group $M$.
The Thompson series $\Tgt$ for $g\in M$ is then defined by
$$
\eqalign{
\Tgt=&\Tr{\HMM}{gq^{L_0}}\cr
     =&{1\over q}+0+(1+\chi_{\rm A}(g))q+...
}
\eqno (2.18)
$$
which depends only on the conjugacy class of $g$ where $\chi_{\rm A}$ is the
character of the 196883 adjoint representation
and where the other coefficients are
similarly positive sums of irreducible characters e.g. for the involution $i$,
$T_i(\tau)=1/ \eta_\r(\tau)+24$.
Likewise, an explicit formula may be found for $g\in C(i\vert M)=
2^{1+24}_+.\Co$ \refs{\FLM, \FLMb} (see \S 4.4).

The Thompson series for the identity element is the partition function
$J(\tau)$.  The compactification
$\overline{\cal F}$ of the
fundamental region ${\cal F}=H/\G$ (where $\G$ is the full modular group
and $H$ is the upper half plane) is isomorphic to the Riemann sphere of genus
zero. The function $J(\tau)$ explicitly realises this isomorphism by
providing a one to one map between $\overline{\cal F}$  and the Riemann sphere.
Such a function is called a hauptmodul for the genus zero modular group $\G$.
A modular invariant meromorphic function is a hauptmodul if and
only if it possess a unique
simple pole on $\overline{\cal F}$.
Once the location of this pole is specified, this function
is itself unique up to constant. Thus $J(\tau)$ is the unique
(up to a constant) modular invariant meromorphic function with
a simple pole at $q=0$ e.g. \refs{\Serre, \Tuiteone}.

Based on \lq experimental\rq\ evidence, Conway and Norton suggested in their
famous paper  \lq Monstrous Moonshine\rq, that the Thompson series
for each $g\in M$ is a hauptmodul (with a simple pole at $q=0$)
 for some genus zero modular
group $\Gg$ under which $\Tgt$ is invariant.  $\Gg$ was explicitly found by
Conway and Norton as follows.

\smallskip

\proclaim Monstrous Moonshine.
Let $g\in M$, $g$  of order  $n$.
   \item{(a)}  The \Th $\Tgt$ is invariant up to $h$ roots of unity under
a subgroup of $\Nor$ of the form
$\Gamma_0(n\vert h)+e_1,e_2,...$  where $h\vert 24$ , $h\vert n$ and $N=nh$.
   \item{(b)}  The subgroup $\Gg$ of these transformations which fixes
$\Tgt$  (and contains $\GN$) is of genus zero where
$\Tgt$  is the corresponding hauptmodul.

 The modular groups $\Gamma_0(n\vert h)+e_1,e_2,...$ and $\Nor$,
the normalizer of $\GN=
\{(\matrix{a& b\cr
              Nc & d\cr})\vert {\rm det }=1\}$
 in ${\rm SL}(2,R)$ are
described in Appendix A.
This result has been rigorously demonstrated by
Borcherds \Borch\ by identifying each Thompson series
with a Weyl-Kac determinant formula for an
associated generalised Kac-Moody algebra.
The proof of Monstrous Moonshine then ultimately relies on a case by case
study of these formulae so that the origin of the genus zero
property remains obscure.
Apart from two classes of order 27, the Thompson series and corresponding
genus zero modular group is unique to each class of
$M$.  Following Conway and Norton, we will abbreviate the notation
denoting the modular groups above and the corresponding Monster group class
in the following way : $\Gamma_0(n\vert h)+e_1,e_2,...$
is abbreviated to $n\vert h+e_1,e_2,...$ and to $n+e_1,e_2,...$ when
$h=1$. If all AL possible involutions are adjoined, these groups are denoted by
$n\vert h+$ and $n+$, respectively, whereas if no AL involutions are
adjoined, then
they are denoted by $n\vert h-$ and $n-$, respectively.
Thus each class of $M$ will be denoted by
$g=n\vert h +e_1,e_2,...$ corresponding
to the modular group for $\Tgt$ in this notation.
As an example,  for the involution $i$, $T_i(\tau)$  is
a hauptmodul for the genus
zero modular group $\G_0(2)$ and $i$ is a member of the class $2-$.

\beginsection {3. Other Constructions of the Moonshine Module}

\subsect{3.1 The FLM uniqueness conjecture.} In the last section we reviewed
the FLM construction of the Moonshine Module $\MM$. There we saw that $\MM$
is a modular invariant meromorphic CFT without any massless states
with partition function $J(\tau)$. FLM have
conjectured that $\MM$ is characterised (up to isomorphism) as
follows \FLMb:

\proclaim  FLM Uniqueness Conjecture.
$\MM$ is the unique meromorphic conformal field theory with
modular invariant partition function $J(\tau)$ and central charge 24.

\noindent
This uniqueness conjecture is analogous to the uniqueness property of the
Leech lattice as being the only even self-dual lattice in 24 dimensions without
roots. In this section we will discuss some evidence to
 support this conjecture by
considering alternative orbifold constructions which are
modular invariant meromorphic
CFTs without massless operators and
with partition function $J(\tau)$. Within these
constructions, we will recognise known properties of the Monster group and will
also find a new relationship between 51 centralisers of the Conway
and Monster groups
generalising an observation made by Conway and Norton \CN.
In the next section we will also link this uniqueness
conjecture to the Monstrous Moonshine properties of Conway and Norton \CN.

\subsect{3.2  $\bf Z_n$ orbifoldings of $\bf \VL$ with \pfu $\bf J(\tau)$.}
Let us now consider orbifold models based on other order $n$
\autos $\{a\}$ of the
untwisted Leech lattice theory $\VL$ \refs{\FLMb,
\Tuitetwo,\DongMason}. $a$ will be chosen so that each model contains no
massless operators, has a meromorphic OPA and is modular invariant
with partition function $J(\tau)$ as in (2.16) and hence, according to the
uniqueness conjecture, reproduces $\MM$. In each construction,
we will also be able to identify an \auto $g_n$ where
$g_n$ (or a power of $g_n$) is \lq dual\rq\ to $a$.
We will find a total of 51 such \autos
which we will argue are representatives of the complete list of  51
Monster group classes $n\vert h+e_1,e_2,...$
with $e_i\not = n/h$ i.e. elements whose Thompson
series are  not invariant under the Fricke involution
$w_n:\tau\rightarrow-1/nh\tau$. Such elements of $M$  are called non-Fricke.
Each stage of the original construction reviewed in \S 2 will be appropriately
generalised  but a rigorous treatment along the lines of FLM is
not yet available in general with the exception of the prime ordered cases
recently described by Dong and Mason \DongMason.

Let us consider an OPA \auto $a$ of $\VL$  lifted from an \auto
$\a\in{\rm Co_0}$ of $\Lambda$ given by
$$
\eqalignno{
a c(\beta)a^{-1}=&e^{2\pi i f_a(\beta)}c(\a\beta)&(3.1a)\cr
a \del_z X^i(z)a^{-1}=&\omega^{s_i}\del_z X^i(z)&(3.1b)\cr
}
$$
where we choose a diagonal basis for
$\a={\rm diag}(\omega^{s_1},...,\omega^{s_{24}})$ with $\omega=e^{2\pi i/n}$.
$f_a(\beta)\in Z/2$ describes the lifting of
$\a$ to an \auto $a$ of $\hat\Lambda$
which preserves (2.6).
We only consider lattice \autos $\a$ without fixed points in
order to ensure that
no untwisted massless states $\del_z X^i(z)$ survive projection under
$\Pa=(1+a+...+a^{n-1})/n$.  This
condition also guarantees that $a$ and $\a$ are of the same order $n$
throughout \Lepowsky.
Each conjugacy class of ${\rm Co_0}$ is parameterised by the
characteristic equation
for a representative element $\a$ as follows
$$
\eqalign{
\det{x-\a}=\prod_{k\vert n}(x^k-1)^{a_k}
}
\eqno (3.2)
$$
 $k\vert n$ denotes that $k$ divides $n$, the order of $\a$ and
each $a_k$ is a not necessarily positive integer where
$$
\eqalign{
 \sum_{k\vert n}ka_k=24, \quad
\sum_{k\vert n}a_k=0}
\eqno{(3.3)}
$$
The absence of fixed points for $\a$ implies the second condition
 and also that $a_1\le 0$.
e.g. $\r$ is parameterised by $r_2=-r_1=24$ with $\det{x-\r}
=(x+1)^{24}$. For $n=p$ prime, the parameters are given by $a_p=-a_1=2d$
where $(p-1)2d=24$ with $d=12,6,3,2,1$ for $p=2,3,5,7,13$.

Since $a$ is an OPA \auto for $\VL$, $\Pa\VL$ also
forms a meromorphic OPA which closes. The associated
partition function $\Tr{\Pa\HL}{q^{L_0}}$ is not
modular invariant, as before,
necessitating the introduction of $b=a^r$ twisted sectors
where $b$ is lifted from
$\b=\a^r$ of order $m=n/(n,r)$ with characteristic equation
parameters $\{b_k\}$.
Thus we find that under the
modular transformation $S:\tau\rightarrow -1/\tau$ \refs{\Narain,\Tuiteone}
$$
\eqalign{
\orbsqr{b}{1}={\Theta_{\L_\b}(\tau)\over \eta_\b(\tau)}\rightarrow
\orbsqr{1}{b}={D_\b^{1/2}\over V_\b}
{\Theta_{\L_\b^*}(\tau)\over \eta_\b^*(\tau)}
={D_\b^{1/2}\over V_\b}q^{E_0^b}(1+O(q^{1/m}))
}
\eqno (3.4)
$$
where
$$
\eqalignno{
\Theta_{\L_\b}(\tau)=&
\sum_{\beta\in \L_\b} q^{\beta^2/2}, \quad
\eta_\b(\tau)=\prod_{k\vert m}\eta(k\tau)^{b_k},\quad
\eta_\b^*(\tau)=\prod_{k\vert m}\eta(\tau/k)^{b_k}&(3.5a)\cr
D_\b=&\prod_{k\vert m}k^{b_k}={\rm det}_T(1-\b), \quad
E_0^b=-{1\over 24}\sum_{k\vert m}{b_k\over k}&(3.5b)\cr}
$$
Here we have chosen the lifting $b$ of $\b$ to an
\auto of $\hat\L$ where $b c(\beta)b^{-1}=c(\beta)$ for $\beta\in\L_\b$,
the sublattice of $\L$  fixed by $\b$.
$\L_\b$ has dual lattice $\L_\b^*=\L_\para \equiv {\cal P}_\b\L_\b$
 and is of volume
$V_\b=\vert \L_\para/\L_\b\vert ^{1/2}$.
The determinant of (3.5b)
denotes the exclusion of all unit eigenvalues of $\b$.
 These expressions simplify
for $b=a$ lifted from $\a$ in which case $\Theta_{\L_\a}(\tau)=1$
and $V_\a=1$.

\subsect{3.3 51 \autos of the Leech lattice.}
We may anticipate some features of a $b$
twisted sector $\Vb$ with the partition
function $\textorbsqr{1}{b}$.
We expect $\Hb$ to have vacuum degeneracy $D_\b^{1/2}/V_\b$
and vacuum energy $E_0^b$. From (3.4), $\textorbsqr{1}{b}$ is
invariant up to a phase $\exp(2\pi i mE_0^b)$ under
 $T^{m}:\tau\rightarrow\tau+m$ i.e.
the action of $b$ on the twisted sector is of order $m$
up to this phase. However, to construct a meromorphic orbifold CFT with a
modular invariant partition function we
must have $mE_0^b=0{\ \rm mod\ }1$ i.e. there is no global phase anomaly
\refs{\Vafa,\FV}. Equivalently, there is no such anomaly
provided $\textorbsqr{b}{1}$ is
invariant under the modular group $\Gamma_0(m)$
\refs{\Vafa,\Tuiteone}.
Lastly, if $E_0^b\le 0$ then the
$b$ twisted sector may reintroduce massless states.
Let us initially consider the $a$ twisted sector here
and study those \autos with $nE_0^a=0\ {\rm mod\ }1$ and $E_0^a>0$
\Tuitetwo. As we will see below, these conditions are
 sufficient to ensure the absence of
a global phase anomaly and massless states in any
of the $b=a^r$ twisted sectors
in the full orbifold construction.
We therefore restrict ourselves to the study of \autos $\a$
obeying \Tuitetwo
$$
\eqalignno{
\sum_{k\vert n}a_k=&0, \quad
E_0^a>0&(3.6a)\cr
nE_0^a=&0\ {\rm mod\ }1&(3.6b)\cr
}
$$
In column 1 of Table 1 we give a complete list of the 38 characteristic classes
of $\rm Co_0$ \Kondo\ that obey the constraints (3.6).
$\a$ with  parameters $a_k,...,a_l,-a_m,...,-a_n>0$ is denoted
by $k^{a_k}...l^{a_l}/m^{a_m}...n^{a_n}$,
called the Frame shape notation. In each case
$a_k$ obeys the symmetry relation
$a_k=-a_{n/k}$ and therefore, from (3.3) and (3.5b), $E_0^a=1/n$.
One may also check that $b=a^r$ of order $m$ obeys
$mE_0^b=0{\rm \ mod\ }1$ and hence no global phase anomaly
occurs in the $b$ twisted sector. Under a general modular tranformation
$\tau\rightarrow (a\tau+b)/(c\tau+d)$ we also find that $\textorbsqr{a}{1}
\rightarrow\textorbsqr{a^d}{a^{-c}}$ in the usual way.
Therefore for $\gamma\in\Gn$ where
$\gamma:\tau\rightarrow (a\tau+b)/(cn\tau+d)$,
$\textorbsqr{a}{1}\rightarrow \textorbsqr{a^d}{1}=\textorbsqr{a}{1}$
since $(d,n)=1$ i.e. $n$ and $d$ are relatively prime and hence
$\eta_{\a^d}=\eta_\a$.
In column 2 we give the full modular invariance group $\Ga$ of
$\textorbsqr{a}{1}$ in the notation described in \S 2.4.
In general, $\Ga$ does not
uniquely specify a class of ${\rm Co_0}$ but does do so for classes obeying
(3.6a). In Table 2 we give a complete list of the remaining
13 classes of $\rm Co_0$
that obey the constraints (3.6a) only. Each of these classes is characterised
by the existence of an integer $h\not=1$  with $h\vert k$ for all $a_k\not =0$
where, from (3.3), $h\vert 24$. In each case the parameters
$\{a_k\}$ obey the symmetry relation $a_k=-a_{nh/k}$ and therefore
$E_0^a=1/nh$ violating (3.6b) for $h\not =1$. Column 2 shows the modular
group $\Ga$ under which $\textorbsqr{a}{1}$ is invariant
up to phases of order $h$ (and hence forms a projective representation of
$\Ga$). This set of classes cannot be employed
to construct a meromorphic orbifold CFT  but is of interest since
for each $\a$ in Table 2, $\a^h$ appears in Table 1.
In general, Table 2 contains all the remaining classes of $\rm Co_0$
with some power in Table 1.

The modular groups $\Ga$ appearing in Tables 1 and 2 are amongst the list
of genus zero groups considered by Conway and Norton \CN\ i.e. for each $\Ga$
there is a corresponding $g_n\in M$ with a Thompson series
$T_{g_n}(\tau)$ of (2.18)
 invariant  under $\Ga$ (up to phases of order $h$).
Furthermore, $\textorbsqr{a}{1}
=1/\eta_\a(\tau)$ is the hauptmodul for $\Ga$ (or, for $h\not =1$, the subgroup
of $\Ga$ that leaves $\textorbsqr{a}{1}$ invariant) and hence
$T_{g_n}(\tau)=1/\eta_\a(\tau)-a_1$
where the constant is fixed by the absence of massless
states in $\MM$. We will identify such an element $g_n$ explicitly below.
Also note that none of these modular groups includes the Fricke involution
$w_n:\tau\rightarrow-1/nh\tau$ since $\eta_\a(\tau)$ is
inverted under $w_n$ with
$\eta_\a(\tau)\rightarrow D^{-1/2}_a/\eta_\a(\tau)$ and
hence $\textorbsqr{1}{a}=
D_\a^{1/2}\eta_\a(\tau/nh)$.  In fact, column 2 of Tables 1 and 2 gives
an exhaustive list of all the modular groups for Thompson series
which are not invariant  under the
Fricke involution i.e the corresponding elements $g_n\in M$ are
the non-Fricke elements.

\subsect{3.4 The $\bf \a$ twisted string construction.}
Let us now consider the construction of the $a$ twisted sector, which is
similar to that of \S 2, for the \autos of both Table 1 and 2.
We will briefly discuss the construction of the
general $b=a^r$ twisted sector later on and in Appendix B.
We introduce $\tilde X^i(z)$ obeying the twisted
monodromy condition $\tilde X^i(e^{2\pi i}z)=
\omega^{-s_i}\tilde X^i(z)+2\pi \beta^i$
(with $\a$ in the diagonal basis) with
mode expansion  \refs{\Lepowsky,\KacP,\Dixon,\CHZn}
$$
\eqalign{
\tilde X^i(z)
=\tilde q^i+i\sum_{m\in Z+s_i/n}{\tilde \alpha^i_m\over m}z^{-m}
}
\eqno (3.7)
$$
where $\tilde\alpha^i_m$ obey the
commutation relations (2.2). $\tilde q^i\in \La=\L/(1-\a)\L$ is the
$\a$ fixed point space of the torus and is a finite
abelian group of order $D_\a = \det {1-\a}$.

The twisted states  $\Ha$ with Virasoro Hamiltonian $L_0=
\sum_m\tilde\alpha^i_m \tilde\alpha^i_{-m}+E_0^a$ and
partition function $\textorbsqr{1}{a}$ of (3.4)
can be again constructed from a set of vertex
\ops $\VLa$ which form a representation of the untwisted set $\VL$.
These act on a degenerate vacuum of dimension $D_\a^{1/2}$
and their OPA forms a
representation of the OPA (2.5) which is a non-meromorphic OPA due
to $Z/n$ grading. The construction of $\tilde\phi(z)\in \VLa$ is similar
to (2.4) where now the cocycle factors are replaced by $\{c_T(\alpha)\}$
defined as follows \refs{\Lepowsky,\KacP}. Consider a
central extension $\Lahat$ of $\La$ by $\langle \omega\rangle $,
the cyclic group generated by $\omega=e^{2\pi i/n}$, given by
$$
\eqalignno{
 c(\alpha) c(\beta) c(\alpha)^{-1} c(\beta)^{-1}=&
{\rm exp}(2\pi i S_a(\alpha,\beta))&(3.8a)\cr
S_a(\alpha,\beta)=-S_a(\beta,\alpha)=
&\langle \alpha,(1-\a)^{-1}\beta\rangle  \ {\rm mod }\ 1
&(3.8b)\cr
}
$$
where $\alpha, \beta$ are representative elements of $\La$
and $S_a(\alpha,\beta)\in Z/n$. Associated with each
section $\{ c(\alpha)\}$ is a 2-cocycle $\epsilon(\alpha,\beta)\in
\langle \omega\rangle $ as in (2.6b) obeying the cocycle condition (2.6c).
In general, the commutator
subgroup of $\Lahat$ is a subgroup of the center $\langle \omega\rangle$
and for $n=p$, prime,
is equivalent to $\langle \omega\rangle$ in which case $\Lahat=p_+^{1+2d}$, an
extra-special $p$ group c.f.\FLMb.  For the full set of \autos obeying (3.6a),
$\Lahat$ is given in column 3 of Tables 1 and 2.
The group $\Lahat$ has a unique irreducible
 faithful representation $\pi$ of dimension $D_\a^{1/2}$ in which the center
is represented by the roots of unity $\langle \omega\rangle
$ \refs{\Lepowsky,\FLMb,\KacP}.
The elements of
$\pi(\Lahat$) are then the cocycle matrices $\{c_T(\alpha)\}$ which act on
a vector space  with basis formed
by the $a$ twisted vacuum states $\{\vert \sigma^l_a\rangle \}$,
$l=1,...D_\a^{1/2}$.

For each \op $\phi(z)\in \VL$ there is a corresponding
\op $\tilde\phi(z)\in \VLa$
which acts on the $a$ twisted vacuum states
$\{\vert \sigma^l_a\rangle \}$ and obeys
the monodromy condition associated with the \auto $a$ as follows
$$
\eqalign{
\tilde\phi^{(k)}(e^{2\pi i}z)=a^{-1}\tilde\phi^{(k)}(z)a=
\omega^{-k}\tilde\phi^{(k)}(z)
}
\eqno(3.9)
$$
where $\phi^{(k)}(z)\in\VL$ is an $\omega^k$ eigenstate of $a$.
The twisted vacuum states are in turn
generated by twist \ops $\{\sigma^l_a(z)\}$
which act on the untwisted vacuum.
For the \autos of Table 1 which
lead to a modular consistent theory, these twist \ops are of conformal
dimension $h_\sigma=1+E_0^a=1+1/n$. The remaining constructions based on the
\autos of Table 2 are discussed below. The construction of $\{\sigma^l_a(z)\}$
can be explicitly performed \CHZn\ where these \ops form a
non-meromorphic OPA with the vertex \ops of $\VL$ and $\VLa$
$$
\eqalign{
\tilde\phi^{(k)}(z)\sigma_a^l(w)=\sigma_a^l(w)\phi^{(k)}(z)\sim
(z-w)^{h_\psi-h_\phi-h_\sigma}\psi^{(k-1)}_a(w)+...
}
\eqno (3.10)
$$
with a suitable analytic continuation assumed in the first equality \CHZn.
$\psi^{(k)}_a(z)$ denotes a conformal field
that creates a twisted state from the untwisted vacuum where
(3.9) implies that the conformal dimension $h_{\psi}\in Z-k/n$. Thus
the first excited twisted states
$\vert \psi^{il}\rangle = \tilde\alpha^i_{-1/n}\vert\sigma_a^l\rangle $
with energy $2/n$ are given by
$\lim_{z\rightarrow 0}z^{(n-1)/n}\del_z\tilde X^i(z)\vert \sigma_a^l\rangle $
for $i=1,...,a_1$
 i.e. they are created by the lowest conformal dimension
\ops $\del_z\tilde X^i(z)$
of $\VLtilde$ which are $\omega^{n-1}$ eigenvectors under $\a$.
We denote the set of \ops  $\{\psi^{(k)}_a(z)\}$, including
 $\{\sigma^l_a(z)\}$, by $\Va$.

The lattice \auto $\a$
 acts as the identity on the fixed point space $\La$. This
allows us to choose a lifting of $\a$ as an \auto of
$\pi(\Lahat)$, which we also denote by $a$, given by
$a c_T(\alpha) a^{-1}=\omega^{-1}c_T(\alpha)$ which is the appropriate choice
for $E_0^a=1/n$. We may then define
the following \auto of the OPA (3.10)
$$
\eqalignno{
a\sigma_a^l(z)a^{-1}=&\omega^{-1}\sigma_a^l(z)&(3.11a)\cr
a\psi^{(k)}_a(z) a^{-1}=
&\omega^k\psi^{(k)}_a(z)=e^{-2\pi ih_\psi}\psi^{(k)}_a(z)&(3.11b)\cr
}
$$
{}From (3.10), the twisted \ops of $\Va$ therefore obey
the twisted mondromy condition
when acting on the vacuum $\vert 0\rangle $
$$
\eqalign{
\psi_a(e^{2\pi i}z)=e^{-2\pi ih_\psi}\psi_a(z)=a\psi_a(z)a^{-1}
}
\eqno(3.12)
$$
Thus $e^{2 \pi i L_0}\vert \psi_a \rangle = a^{-1} \vert \psi_a\rangle$
which implies that under $T:\tau\rightarrow\tau+1$, $\textorbsqr{1}{a}
\rightarrow \textorbsqr{a^{-1}}{a}$ in the expected way e.g. \Ginsparg.
The lifting of $\a$ chosen therefore ensures that no extra phase occurs in this
transformation and that there is no global phase anomaly \refs{\Vafa,\FV}.

For the \autos of Table 2, the twist \ops have conformal dimension
$h_\sigma=1+1/nh$ and $\psi_a^{(k)}(z)$ has conformal
dimension $h_{\psi^{(k)}}\in Z-(k+1)/n+1/nh$.  (3.11) must therefore be
modified
where now $\sigma_a^l(z)$ and $\psi_a^{(k)}(z)$ are, respectively,
unit and $\omega^{k+1}=\omega^{1/h}e^{-2\pi i h_\psi}$ eigenstates under $a$.
Likewise, an extra phase of $\omega^{-1/h}$ appears on the RHS of (3.12).
Hence $\textorbsqr{1}{a}$ is invariant under $T^n$ only up to an overall
global phase of $e^{2\pi i/h}$ giving the
global phase anomaly anticipated earlier.
Furthermore,  from (3.10), the twisted \ops of $\Va$ do not form a
meromorphic OPA with respect to $\Pa\VL$ and hence a meromorphic
orbifold CFT is impossible to construct in these cases.

Examining the twisted partition function
for these cases, we also notice that it is
related to that for $a^h$ with $\textorbsqr{1}{a^h}(\tau)=
[\textorbsqr{1}{a}(h\tau)]^h$ where $D_{\a^h}=D_\a^h$ and
$\eta_{\a^h}(\tau)=[\eta_\a(\tau/h)]^h$ in (3.4). This observation leads us to
an isomorphism between the corresponding twisted Hilbert spaces with
$$
\eqalign{
{\cal H}_{a^h} \cong \Ha\otimes...\otimes\Ha
}
\eqno(3.13)
$$
where the RHS denotes a tensor product over $h$ copies of $\Ha$.
The explicit form of this isomorphism is found by first
 noting that $L_{\a^h}\cong \La\times... \times\La$
for each \auto $\a$ of Table 2.
Since $\a^h$ has no fixed points we have $(1-\a)^{-1}=
(1-\a^h)^{-1}(1+\a+...+\a^{h-1})$ so that  the commutator
subgroup of $\Lahat$ obeys $[\Lahat,\Lahat]\subseteq \langle \omega^h\rangle$
from (3.8b).
The representation $\pi(\Lahat)$ acts on a vector space $T^\a$ of dimension
$D_a^{1/2}$ where the centre is represented by the cyclic group of phases
$\langle \omega\rangle$. Thus $T^\a$ defines the vector space for a
projective representation for $L_\a$ with phases in  $\langle \omega^h\rangle$.
Taking the tensor product of $h$ copies of $T^\a$ we obtain the
vector space $T^\a\otimes...\otimes T^\a$ for the representation
$\pi(\hat L_{\a^h})$ of dimension $D_{a^h}^{1/2}=D_\a^{h/2}$ which forms a
projective representation for  $L_{\a^h}\cong \La\times... \times\La$
with phases in  $\langle \omega^h\rangle$.
Thus the vacuum states of the
twisted Hilbert spaces of (3.13) are isomorphic. Now define
$\tilde\Phi_{i_1...i_h}(z)=
\tilde\phi_{i_1}(z^h)\otimes...\otimes \tilde\phi_{i_h}(z^h)$
which acts on these
twisted vacuum states. Then $\tilde\Phi_{i_1...i_h}(z)$ obeys the
monodromy condition (3.9) for $a^h$. The \ops $\{ \Phi(z)\}$ obey a
non-meromorphic OPA due to $hZ/n$ grading and create Virasoro
eigenstates in ${\cal H}_{a^h }$ (but
are not primary conformal fields in $\Vah$).
The vacuum states of ${\cal H}_{a^h}$ which are created by the twist
\ops $\Sigma_{a^h}^{l_1...l_h}(z)=
\sigma_a^{l_1}(z^h)\otimes...\otimes \sigma_a^{l_h}(z^h)$
 have energy $h_\Sigma=h/n$ and hence the global phase anomaly disappears by
taking this tensor product. Thus the isomorphism between Hilbert spaces  in
(3.13) follows.

We may repeat the $\Va$ construction above for the remaining sectors $\Vb$ with
$X^i(z)$ twisted by $\b=\a^r$ in (3.7b). This is briefly reviewed in Appendix
B.
For $r$ relatively prime to $n$, $\b$ is of order
$n$ also and $\Vb$ is isomorphic
to $\Va$. Otherwise, $\b$ may have unit
eigenvalues and (3.7b) must be modified to
include a momentum component belonging to
$\L_\para$ and where now $\tilde q^i$ lies in the $\b$ fixed point space of the
torus $\Lb=\L_\b^T/(1-\b)\L_T$ with $\L_\b^T=\{\beta\in \L\vert
{\cal P}_\b\beta=0\}$, $\L_T=(1-{\cal P}_\b)\L$.
$\Lb$ is a finite abelian group of order
$D_\b/V_\b^2$.
The construction of the $D_\b^{1/2}/V_\b$ twisted vacuum states
$\{\vert\sigma_b^l\rangle \}$
can be similarly defined \refs{\Lepowsky,\CHZn} together with vertex \ops
$\VLb$ which create $\Hb$ with partition function $\textorbsqr{1}{b}$ of (3.4).
Likewise, the non-meromorphic OPA of (3.10) and mondromy conditions of
(3.9) and  (3.12) are generalised with $a$ replaced by $b$ throughout and
$\Va$ replaced by $\Vb$. These other twisted sectors are
required for modular invariance and for
the expected closure of the corresponding meromorphic OPA.
In particular, we expect the original $a$
twisted \ops $\{\sigma_a^l(z)\}$ to form an
intertwining non-local OPA with the \ops of each sector where
$$
\eqalign{
\tilde\psi^{(k)}_{b}(z)\sigma_a(w)\sim
(z-w)^{h_\chi-h_\psi-h_\sigma}\chi_{ab}^{(k-1)}(w)+...
}
\eqno(3.14)
$$
where  $\psi_{b}^{(k)}\in {\cal V}_{b},\ \chi_{ab}^{(k)}\in {\cal V}_{ab}$
are $\omega^k $ eigenstates of $a$ and where for each $\psi_b(z)\in \Vb$, there
is an \op $\tilde\psi_b(z)$ which acts on the
$a$ twisted vacuum creating a state
in the $ab$ twisted sector. The $b=a^r$ monodromy condition (generalised from
(3.12)) implies that $\psi_{b}^{(k)}$ has conformal dimension
$h_\psi\in Z-kr/n$.

We therefore enlarge the meromorphic set of \ops $\VL$ by
the introduction of the twisted \ops $\{\sigma^l_a\}$  to the set of \ops
$\Vp=\VL\oplus\Va\oplus...\oplus {\cal V}_{a^{n-1}}$ which forms a closed but
non-meromorphic OPA. $\Vp$ consists of all \ops which form a meromorphic
OPA with $\Pa\VL$ i.e. $\Vp$ and $\Pa\VL$ are dual \Goddard. Then
$\Vorb=\Pa\Vp$ forms a closed meromorphic OPA which is self-dual.
Note that only this meromorphic
$\Pa$ projection of the intertwining OPA (3.14) has been rigorously constructed
 and then only in the prime ordered cases $p=2$ in
\refs{\FLMzero, \CHZtwo, \DGMZtwo} and for $p=3,5,7,13$ in \DongMason.
We will assume that (3.14) is true in general.
The \pfu for the corresponding space of states
$\Horb$ is modular invariant with
a unique simple pole at $q=0$ as before and is therefore given by
$Z_{\rm orb}(\tau)=J(\tau)+N_0$ where $N_0$ is the number of massless
operators.
The condition $E_0^a>0$ ensures
that no massless \ops occur in the $a$ twisted sector i.e.
there is no $a$ invariant \op $\psi^{(0)}(z)$ with $h_\psi=1$ which satisfies a
meromorphic monodromy condition $\psi^{(0)} (e^{2\pi i}z)=\psi^{(0)}(z)$ from
(3.12). Nevertheless, there
may be a massless \op $\psi^{(0)}(z)$ present in one of the other $b=a^r$
twisted sectors where $\psi^{(0)}(z)$ is $b$ invariant from the $b$ monodromy
condition (e.g. for $\a=4^8/1^8$, the twisted sector corresponding to
$\b=\a^2=2^{16}/1^8$ has a massless vacuum from (3.5b)).
Taking the $a$ invariant projection
we find $\Pa\psi^{(0)}=0$ unless $\psi^{(0)}(z)$ is also $a$ invariant and
therefore  contradicts our assumption. Thus no massless \ops
that may occur in the other twisted sectors can survive the $\Pa$ projection
and
hence the condition $E_0^a>0$ is sufficient to
ensure the absence of massless \ops in $\Vorb$ and the \pfu is
$Z_{\rm orb}(\tau)=J(\tau)$ once again.  Therefore, according to the FLM
uniqueness conjecture, we  expect $\Vorb\equiv\MM$ for each of the 38 \autos
of Table 1. Let us now consider some evidence to support this.

\subsect{3.5  Centralisers, Thompson series and
$\bf Z_n$ reorbifolding $\bf \Vorb$.}
Let $\Morb$ be the \auto group of the OPA for $\Vorb$ which, from the
FLM uniqueness conjecture, we expect to be $M$, the Monster group.
For the prime ordered cases $p=3,5,7$ and 13, Dong and Mason have
recently demonstrated that $\Morb\equiv M$ for
$p=3$ and very nearly so for $p=5,7,13$ \DongMason.
We may identify an \auto $a^*\in \Morb$ of order $n$
(which generalises the fermion number involution $i$ in the original FLM
construction) under which  the \ops of $\Vak$ are eigenvectors with eigenvalue
$\omega^{k}$. From (3.14), $a^*$ is also an \auto of the
non-meromorphic OPA for the enlarged set of \ops
$\Vp=\VL\oplus\Va\oplus...\oplus {\cal V}_{a^{n-1}}$ and $a^*$ is
 \lq dual\rq\ to the \auto $a$ i.e. the $a$ invariant subset of $\Vp$
is $\Vorb$ whereas the  $a^*$ invariant subset is $\VL$.
Furthermore, $\Vp$ is the set of all \ops
which form a meromorphic OPA with  $\Pa\VL={\cal P}_{a^*}\Vorb$
and hence we may reorbifold $\Vorb$ with
respect to $a^*$ to reproduce $\VL$.
We can see this explicitly as follows. Consider the massless states
$\{\alpha^i_{-1}\vert 0\rangle \}, \ i=1,...-a_1$, which are  $\omega$
eigenstates of $a$.
The \ops of $\Vorb$ obey the $a^*$ twisted monodromy condition when acting
on these states :
$$
\eqalign{
\psi_b^{(0)}(e^{2\pi i}z)=\omega^{-r}\psi_b^{(0)}(z)
={a^*}^{-1}\psi_b^{(0)} (z)a^*
}
\eqno(3.15)
$$
which is analogous to (3.9) i.e. the $-a_1$ massless \ops
$\{\del_z X^i(z)\}, \ i=1,...-a_1$ implement the $a^*$ monodromy condition
for $\Vorb$ and create the $a^*$ twisted vacuum states. The
resulting non-meromorphic OPA closes once again in the enlarged set
$\Vp$ of which the $a^*$ invariant subset is $\VL$. Thus
$$
\eqalign{
\matrix{
 \Vp  \cr
{\buildrel {\cal P}_{a^*}\ \ \over\swarrow} \qquad
 {\buildrel\ \  \Pa\over\searrow}  \cr
 \VL\quad
         \matrix{{\buildrel a\over \longrightarrow}  \cr
                {\buildrel a^*\over \longleftarrow}}
\quad\Vorb\cr}
}
\eqno (3.16)
$$
where the horizontal(diagonal) arrows represent orbifolding(projecting)
with respect to the denoted \aut.

We may also compute the Thompson series $\Torb{a^*}$
for $a^*\in \Morb$ by taking the trace
over $\Horb$, the Hilbert space of states created by $\Vorb$, as follows :
$$
\eqalign{
\Torb{a^*} =\Tr{\Horb}{a^* q^{L_0}}=
\orbsqr{\Pa}{1}+\omega\orbsqr{\Pa}{a}+...
+\omega^{n-1}\orbsqr{\Pa}{a^{n-1}}
}
\eqno(3.17)
$$
For $n=p$, prime, $a^*$ is of prime order and hence
$\sum_{k=1}^p \Torb{a^{*k}}=J(\tau)+(p-1)\Torb{a^{*}}$. This is also equal to
$\textorbsqr{1}{1}+(p-1)\textorbsqr{a}{1}$ from (3.17) where $\Sigma_k {a^*}^k$
vanishes on each twisted sector. Therefore we find that
$\Torb{a^*}=\textorbsqr{a}{1}+24/(p-1)=1/\eta_\a(\tau)+2d$. Thus $a^*\in \Morb$
has the same Thompson series as $p-\in M$ with genus zero modular group $\Gp$.
We can  show that this  generalizes to all orbifoldings
generated by the elements of Table 1 where
$$
\eqalign{
\Torb{a^*}={1 \over \eta_\a(\tau)}-a_1
}
\eqno(3.18)
$$
which is the hauptmodul for the genus zero modular group $n+e_1,e_2,...$ .
This result follows from a  consideration of  the singularities of $\Torb{a^*}$
and showing that they agree with those of $1/\eta_\a(\tau)$ \Tuitetwo.
Thus each $a^*\in \Morb$, the \auto of $\Vorb$ dual to $a$, has the same
Thompson series as the non-Fricke elements $n+e_1,e_2,...\in M$, $e_i\not =n$.

(3.18) may be  generalized to include the other \autos $\{\a\}$ of Table 2.
As already described, such \autos cannot be used to construct a
meromorphic orbifold CFT. However, $\a'=\a^h$, of order $m=n/h$, can
be so employed to construct an orbifold with partition function $J(\tau)$. Let
$g_n$ denote the lifting of $\a$ where $g_n^h=a'^*$ is dual to $a'=a^h$, a
lifting of
$\a'$.  $g_n$ then acts on each twisted space and is in the centraliser of
$a'^*$ in $\Morbh$ (see below). We may compute the
Thompson series for $g_n$ as a trace over
${\cal H}_{\rm orb}^{a^h}$ by a similar  trick to the prime ordered cases
above.
$g_n^{1+hk}=g_n {a'^*}^k$  is of order
$n$ for each $k=1,2,...m$ and  has the same Thompson series as $g_n$.
 Likewise, for each $k$,
$\textorbsqr{g_n}{1}=\textorbsqr{g_n^{1+hk}}{1}$ and therefore
$T_{g_n}(\tau)={1\over m}\sum_k T_{g_n^{1+hk}}(\tau)=\textorbsqr{g_n}{1}=
1/\eta_\a(\tau)$ where $\sum_k {a'^*}^k$ vanishes on each twisted sector.
Thus (3.18) also holds for the \auto $g_n$ (since
$a_1=0$ for $h\not =1$) and $g_n$ has the same Thompson series as
$n\vert h+e_1,e_2,...$, with $e_i \not =n/h$ and $h\not =1$, a
non-Fricke element.

We may next compute the centraliser
$C(g_n\vert \Morbh)=\{g\in \Morbh\vert g_n^{-1}gg_n=g\}$. For the
 38 \autos with
$h=1$ this consists of all OPA \autos that do not mix the various projected
sectors $\Pa\Vak$ of $\Vorb$.  For the remaining 13 \autos $g_n$ with
$h\not =1$,  $C(g_n\vert \Morbh)\subset
C(a'^*\vert \Morbh)$. Every element $g\in C(a^*\vert \Morb)$ must
commute with $a$ in order to preserve the $\Pa$ projection.
Thus $C(a^*\vert \Morb)$ is some extension of
$G_n=C(\a\vert{\rm Co}_0)/\langle \a\rangle $, the non-trivial part
of the Conway group centraliser, which is reproduced from \Wilson\ in column 4
of Tables 1 and 2.
The nature of this extension can be seen by considering the \auto group
preserving  the twisted sector $\Pa\Va$ \Tuitetwo.
Let $g$ and $g'$ be two inequivalent liftings of $\g$ to \autos of
$\pi(\Lahat)$, the
faithful representation of $\Lahat$ whose elements are the $a$ twisted cocycle
matrices $\{c_T(\alpha)\}$. Thus $g'g^{-1}$ is a lifting of the
identity lattice
\aut. However, the
inner \autos of $\pi(\Lahat)$ given by $c_T(\alpha):c_T(\beta)\rightarrow
c_T(\alpha)c_T(\beta)c_T(\alpha)^{-1}=exp(2\pi iS_\a(\alpha,\beta))c_T(\beta)$
describe the inequivalent liftings of the identity and hence the inequivalent
liftings $g$ of $\g$. As discussed above
in (3.11), the lifting $a$ of $\a$ to an \auto of
$\pi(\Lahat)$ is $ac_T(\alpha)a^{-1}=\omega^{-1}c_T(\alpha)$. Hence
$g$ commutes with $a$ and in turn, defines an \auto for
$\Pa\Va$ through (3.10). Thus we find that the group of inequivalent OPA \autos
preserving $\Pa\Va$ is $\Lahat.G_n$, an extension of $G_n$.
The same result also holds for the isomorphic twisted sectors $\Pa\Vak$
where $a^k$ is of order $n$ i.e. $k$ is relatively prime to $n$.
In Appendix B we discuss the contribution of the remaining sectors to
$C(a^*\vert \Morb)$. There we also consider the other 13 \autos with
$h\not =1$ and demonstrate that for all 51 \autos $g_n$
$$
\eqalign{
C(g_n\vert \Morbh)=\Lahat.G_n}
\eqno(3.19)
$$
In column 5 of  Tables 1 and 2 we have reproduced $C(g_n\vert M)$ from
\CN\ which may be compared with $\Lahat$ and $G_n$ in columns 3 and 4
to verify (3.19) assuming that $\Morb\equiv M$ and
$g_n\equiv n\vert h+e_1,e_2,...$, a non-Fricke element of $M$.
(3.19) is a new generalisation of the original observation
of Conway and Norton concerning the five $n=p$, prime, cases where
$C(p-\vert M)=p_+^{1+2d}.G_p$ with $a^*=p-$ \CN .
For the other 46 \autos of Tables 1 and 2, there are only 11 cases for which
(3.19) can be explicitly checked using the
available information about these centralisers in \refs{\CN,\Wilson}. However,
the order of these groups agrees with (3.19) in each case supporting the
very likely validity of the result in general.

{}From (3.10) we may observe that $\Lahat.G_n$ must be an extension of
$\hat G_n=C(a\vert 2^{24}.{\rm Co}_0)/\langle a\rangle $,
the subgroup of \autos of $\VL$
which preserve $\Pa\VL$, where the extension contains the central cyclic group
generated by $g_n$. This extension is due to the presence of the
$D_\a^{1/2}$ twist \ops $\{ \sigma^l_a\}$ which form a representation of
$\hat G_n$. Thus for the prime ordered cases  $\hat G_2=2^{24}.\Co$ and
$\hat G_p=G_p$ for $p=3,5,7,13$. In particular, we also note that
if the $a$ twisted vacuum is unique, then $\Lahat.G_n$ is isomorphic to
$n.\hat G_n$. A similar observation will be
useful in  \S 4 when we consider other possible orbifoldings of $\MM$.

\subsect{3.6  A $\bf Z_2$ reorbifolding of $\bf \Vorb$.}
Recently, Montague  made the interesting suggestion \Mont\ that a CFT,
such as $\Vorb$, with \pfu $J(\tau)$ can be shown to be isomorphic to $\MM$ by
the existence of an involution $i$ of $\Vorb$ and a set of
twisted \ops ${\cal V}_i$ with non-negative vacuum energy (see \S 4). Then
the Thompson series $\Torb{i}$ is $\G_0(2)$ invariant with a unique simple pole
at $q=0$ and must be the hauptmodul $1/\eta_\r(\tau)+24$. Therefore,
assuming that we can reorbifold $\Vorb$ with respect to $i$,
we obtain  a CFT with \pfu $J(\tau)+24$.
But $\VL$ is now known to be the unique CFT with this \pfu
\Mont\ and hence this reorbifolding
reproduces $\VL$. If we consider the involution $i^*$ dual to $i$
 which acts on $\VL$, then
the 24 massless \ops of $\VL$ are $-1$ eigenvectors under $i^*$ and hence
$i^*$ can be identified with the involution $r$ introduced in the original FLM
construction. Thus $\Vorb$ can be  obtained from $\VL$ by orbifolding
with respect to $i^*\equiv r$ and must be isomorphic to $\MM$.

We will now consider the constructions of $\Vorb$ given above and find
an involution $i$ with the correct Thompson
series in 11 cases in addition to the
standard FLM construction.
We will only consider here an involution in the
centraliser $C(a^*\vert \Morb)=\Lahat.G_n$ which is lifted from the reflection
\auto $\r$ of $\L$.  This restriction excludes 13  \autos of
even  order $n$  (including the original \auto $\r$ adopted by FLM !)
denoted by $\dagger$ in the last column of  Table 1
for which $\a^{n/2}=\r=-1$ so that $\r \not\in G_n$. For the remaining
\autos we can compute $\Torb{i}$ similarly to (3.17).
Under $S:\tau\rightarrow-1/\tau$ we obtain given the usual modular
transformation properties
$$
\eqalign{
T^{\rm orb}_i(-{1\over \tau})=
\orbsqr{\Pa}{i}+\orbsqr{\Pa}{ia}+...+\orbsqr{\Pa}{ia^{n-1}}}
\eqno (3.20)
$$
which is $T^2$ invariant and hence $\Torb{i}$ is $\G_0(2)$ invariant.
We can determine whether $\Torb{i}$ is a hauptmodul for $\G_0(2)$
 by considering the behaviour at $\tau=0$ via (3.20).
The sector twisted by $i$ has vacuum energy  $+1/2$ because $i$
is lifted from $\r$ and therefore contributes no singularity.  Each
sector twisted by $ia^k$, of order $m$,  has vacuum energy  which always obeys
$E_0\ge -1/m$ (see \S 4.4) and
therefore contributes no singularity unless $m=2$ with $E_0=-1/2$
since (3.20) is $T^2$ invariant.
This occurs when $ia^k$ is lifted from
$-\a^k$ with Frame shape  $1^8.2^8$ i.e.
$\a$ is of even order $n=2k$ and
$\a^k$ has Frame shape $2^{16}/1^8$ which is the
case for the 14 \autos denoted by  $\ddagger$ in Table 1.
Otherwise, for the 11 remaining \auts, denoted by  $*$ in
Table 1,  $\Torb{i}$ has a unique simple pole
at $q=0$ and is therefore a hauptmodul
for $\G_0(2)$. These consist of 3 even ordered \autos and all the odd ordered
\autos including the odd prime ones considered by Dong and Mason \DongMason.
Thus, in  these 11 cases, one can construct the required involution.
In the remaining
cases, a more technical construction is required and is currently under
investigation.

To summarise this section, we have described 38 meromorphic
orbifold constructions $\Vorb$ (including the original one of FLM and the
prime ordered constructions of Dong and Mason) with partition
function $J(\tau)$. Amongst these constructions, we
have found 51 \autos $\{g_n\}$
that can be identified with the 51 non-Fricke Monster group classes where
$g_n$ satisfies $g_n^h=a'^*$, the \auto dual to $a'=a^h$. For each $g_n$, the
Thompson series  agrees with the corresponding Monster
group Thompson series and the centraliser in (3.19) also agrees explicitly in
many cases (and very probably in all cases). For 11 of these new constructions,
an involution can also be found which is dual to the
involution $r$ of $\VL$ used in the FLM
construction of $\MM$ and so $\Vorb\equiv\MM$ for
these cases (assuming that the
various twisted sectors obey the OPAs (3.10) and (3.14)).
We also note that we may in general compute the
Thompson series within $\Vorb$ for each element of $C(g_n\vert \Morb)$ as a sum
of traces over each sector $\Pa\Hb$ (in \Tuitetwo\ we give an explicit formula
for the prime ordered constructions). In particular, it
is straightforward to show
 that $T_{g_n^k}(\tau)$ agrees with the expected result in each case.
All of these results support the conjecture that
$\Vorb\equiv \MM$ as expected from the FLM uniqueness conjecture. Finally,
we expect a generalised version of the
hidden triality symmetry in the FLM construction which mixes the untwisted and
twisted sectors to exist \refs{\FLMzero,\FLMb,\DGMtri}.
Thus there should exist some symmetry group
$\Sigma_n$ which mixes the various sectors
of $\Vorb$ where $C(g_n\vert M)$ and $\Sigma_n$ generate $M$. In the
prime cases $p=3,5,7,13$, $\Sigma_p$ has been constructed by Dong and Mason
\DongMason.

\beginsection {4. Orbifolding the Moonshine Module and Monstrous Moonshine}

\subsect {4.1 Monstrous Moonshine and orbifolding $\bf \MM$. }
Let us now consider one of the the main objectives of this
paper which is to discuss the relationship of the FLM uniqueness conjecture
to Monstrous Moonshine, the genus zero property for Thompson series \CN.
Our main result is as follows: Assuming the FLM uniqueness
conjecture holds, then
the Thompson series for $g\in M$ is a hauptmodul if and only if the
only meromorphic orbifoldings of $\MM$
with respect to $g$ are  $\VL$ or   $\MM$.

We will assume throughout this section that the FLM
uniqueness conjecture is correct.
Therefore $\Vorb\cong\MM$ for each of the
orbifoldings described in \S 3 and $\VL$ can be reconstructed by reorbifolding
$\MM$ with respect to the non-Fricke dual \autos  ${a^*}=n+e_1,e_2,...$ with
$e_i\not = n$. The Thompson series
for ${a^*}$ of (3.15) is then recognised as a
contribution to the partition function for this reorbifolding.
It is natural to interprete all the Thompson series $\Tgt$
 in this way and to construct an orbifolding
of $\MM$ with respect to each $g\in M$ \Tuiteone. In particular, we
expect that under $S:\tau\rightarrow-1/\tau$, $T_g(\tau)=\Tr{\HMM}{gq^{L_0}}$
transforms to the partition function for a $g$ twisted sector as follows:
$$
\eqalign{
T_g(\tau)=\MMorbsqr{g}{1}\rightarrow \MMorbsqr{1}{g}+...
}
\eqno (4.1)
$$
where the superscript $\natural$ denotes
a trace contribution to the orbifolding
of $\MM$  (in distinction to orbifoldings of $\VL$) and where
the $g$ twisted sector $\Vg$ has vacuum energy $E_0^g$ and
degeneracy $N_g$. For the 38  \autos ${a^*}$
dual to $a$ we find from (3.16) that
$\MMtextorbsqr{1}{{a^*}}=-a_1+D_\a^{1/2}\eta_\a(\tau/n)$ with vacuum energy
$E_0^{{a^*}}=0$ and degeneracy $N_{{a^*}}=-a_1$. In these cases,
${\cal V}_{{a^*}}^\natural=\{\phi^{(1)}\}\oplus \{\psi^{(1)}_a\}\oplus...
\oplus \{\psi^{(1)}_{a^{n-1}}\}$, the subspace of
$\VL\oplus\Va\oplus...\oplus{\cal V}_{a^{n-1}}$
with eigenvalue $\omega$ under
$a$ where, as noted in \S 3, the ${a^*}$ twisted vacuum is created
by the massless
\ops $\del_z X^i(z),\ i=1,...-a_1$. Likewise, the other 13 non-Fricke
\autos $g_n$ with $g_n^h={b^*}$
(where $b^*$ is dual to $a^h$ and $h \not =1$)
have vacuum energy $E_0^{g_n}=1/nh$ and degeneracy
$N_{g_n}=D_\a^{1/2}$ and therefore possesses a global phase anomaly leading to
an orbifold construction which is not meromorphic and
not consistent with modular symmetry
\refs{\Vafa,\FV}.
The twisted space of \ops ${\cal V}_{g_n}^\natural$ will be discussed in
\S 4.3 and \S 4.4 below.
For the remaining Fricke classes of $M$, $g=n\vert h+{n\over h},e_2,...$
(i.e. $e_1={n\over h}$), we will assume that the twisted \op sector
$\Vg$, with a corresponding Hilbert space of states $\Hg$, can always be
constructed. There are a total of
120 of these classes (including two classes 27A, 27B which have the same
Thompson series) of which 82 classes have $h=1$ \CN. For many of these
classes, the method of construction of these
sectors is not known since the origin
of the \auto is not geometrical as was the case for the lattice \autos of \S 3.
However, for \autos in the centraliser $C(i\vert M)=2^{1+24}.\Co$
which are associated
with Leech lattice \auts, a method of construction is given later on \S 4.4.

The $q^n$ coefficients of the trace on the RHS of (4.1)
must all be non-negative since this is the partition function
$\Tr{\Hg}{q^{L_0}}$ for the Hilbert space $\Hg$ associated with $\Vg$.
(In fact, from the point of view of the representation theory of
Virasoro algebras,  $\Tr{\Hg}{q^{L_0}}$ is the
characteristic function and is arguably a more natural
object to study than the original Thompson series).
For the Fricke classes $\Tgt=\MMtextorbsqr{1}{g}(nh\tau)$ whereas
for the non-Fricke classes
$T_{g_n}(\tau)=1/\eta_\a(\tau)-a_1
=-a_1+D_\a^{1/2}/(a_1+\MMtextorbsqr{1}{g_n}(nh\tau))$. Therefore the
$q^n$ coefficients of $\Tgt$ must be non-negative for the Fricke
classes and of mixed sign for the non-Fricke classes.  These
properties are  indeed observed for all Thompson series.

For orbifold constructions leading to a theory with modular consistency,
the vacuum energy $E_0^g$ must also satisfy
$nE_0^g=0\ {\rm mod}\ 1$ and
$\MMtextorbsqr{1}{g}$ is $T^n$ invariant. Assuming the usual orbifold
trace modular transformation properties, for all
$\gamma\in\Gn$ where $\gamma:\tau\rightarrow (a\tau+b)/(cn\tau+d)$ we
 find  $\Tgt\rightarrow \MMtextorbsqr{g^d}{1}=\Tgt$ since $(d,n)=1$ i.e.
$n$ and $d$ are relatively prime so that $g$ and  $g^d$ are
in the same conjugacy class
and hence have the same Thompson series. Thus, in the absence of a global
phase anomaly, $\Tgt$ is $\Gn$ invariant and hence $h=1$.
Let us consider, for the present, only Thompson series with this property.

In general, we assume that there exists a set of \ops $\{\sigma_g^l(z)\}$,
$l=1,...,N_g$ of conformal dimension $h_\sigma=1+j/n$ which create
the  vacuum \ops of $\Vg$. We also
assume that for each \op $\psi(z)\in\MM$, there is an operator
$\tilde\psi(z)$, which acts on this twisted vacuum and creates a state in
$\Hg$.
If $\psi^{(k)}(z)\in\MM$ is an $\omega^k$ eigenstate of $g$
then we assume that when acting on
the vacuum states $\{\vert\sigma_g^l\rangle \}$,
$\tilde\psi^{(k)}(z)$ satisfies the following monodromy condition :
$$
\eqalign{
\tilde\psi^{(k)}(e^{2\pi i}z)=\omega^{-k}\tilde\psi^{(k)}(z)=
g^{-1}\tilde\psi^{(k)}(z)g
}
\eqno(4.2)
$$
Similarly to (3.9) and (3.10), (4.2) follows
from a non-meromorphic OPA which the
twisted \ops $\{\sigma_g^l(z)\}$ satisfy with $\MM$ where
 $$
\eqalign{
\tilde\psi^{(k)}(z)\sigma_g^l(w)=\sigma_g^l(w)\psi^{(k)}(z)\sim
(z-w)^{h_\chi-h_\psi-h_\sigma}\chi_g^{(k-j)}(w)...\cr
}
\eqno(4.3)
$$
where the \ops $\{\sigma_g^l(z)\}$ are $\omega^{-j}$
eigenvectors of $g$ and $\chi_g^{(k)}(z)\in \Vg$ has conformal dimension
$h_\chi\in Z-k/n$  and is an $\omega^k$ eigenvector of $g$.
Then each $\chi_g(z)\in \Vg$ obeys the usual mondromy
condition
$$
\eqalign{
\chi_g(e^{2\pi i}z)=g\chi_g(z)g^{-1}=e^{-2\pi i h_\chi}\chi_g(z)
}
\eqno(4.4)
$$
when  acting on the untwisted vacuum $\vert 0\rangle $ so that
$T:\MMtextorbsqr{1}{g}\rightarrow\MMtextorbsqr{g^{-1}}{g}$ as expected,
without any global phase anomaly. Likewise, the twisted sectors
$\{\Vgk\}$ are assumed to exist with vacuum energy $E_0^{g^k}$ and degeneracy
$N_{g^k}$ where together
$\Vp=\MM\oplus\Vg \oplus ...\oplus {\cal V}_{g^{n-1}}^\natural$
forms a closed non-meromorphic OPA. Taking the projection we define
$\MMorb= \Pg\Vp$,
the CFT constructed from $\MM$ by orbifolding with respect to $g$. The \ops
of $\MMorb$ form a meromorphic OPA and the partition function is again
$Z_{\rm orb}(\tau)=J(\tau)+N_0$ where $N_0$ is the number of massless
operators.
For each of the 38 non-Fricke \autos ${a^*}$ dual to $a$, this
construction give us $\VL$ with $N_0=24$.  Assuming that Thompson series
are hauptmoduls, we will show below that $N_0=0$ for the remaining
82 global phase anomaly free Fricke classes (which we denote by
$f=n+n,e_2,...$)   so that $\MMforb\cong \MM$ again i.e. every meromorphic
orbifolding of $\MM$ with respect to  $g\in M$ either produces
$\VL$ or reproduces $\MM$ again. Conversely, we will  show in
\S 4.2 that given this
result then $\Tgt$  must be a hauptmodul  for some genus zero modular group.

We begin by describing how  $\Tgt$ can be a  hauptmodul in terms
of the vacuum properties of  $\Vgk$ for a meromorphic
 orbifolding of $\MM$ with respect to $g$.
We assume that under a general
modular transformation $\gamma(\tau)=(a\tau+b)/(c\tau+d)$,
${T_g}(\gamma(\tau))=\MMtextorbsqr{g^{-d}}{g^c}$.
Thus any possible singular behaviour of $\Tgt$ at a cusp point
$a/c=\lim_{\tau\rightarrow\infty}\gamma(\tau)$ is governed by the
vacuum energy and degeneracy of the $g^c$ twisted sector.
In \Tuiteone\ we showed that for
$g=n+e_1,e_2,...\in M$, $\Tgt=\MMtextorbsqr{g}{1}$ is a hauptmodul for the
modular group $\Gg=\Gn+e_1,e_2,...$ if and only if the vacuum energies and
degeneracies of the twisted sectors $\Vgk$ obey the following properties

\proclaim Vacuum Properties.
\smallskip
{(I)} The vacuum energy $E_0^{g^k}$ for $\Vgk$ is non-negative unless $g^k$ is
of order $e\in\{e_1,e_2,... \}$ in which case $E_0^{g^k}=-1/e$ ($\Vgk$ is
tachyonic) and the vacuum degeneracy $N_{g^k}=1$.
\smallskip
{(II)} (Atkin-Lehner Closure) If both sectors ${\cal V}_{g^{k_1}}^\natural$ and
${\cal V}_{g^{k_2}}^\natural$ are tachyonic (with vacuum
energies $-1/e_1,-1/e_2$) then the sector ${\cal V}_{g^{k_3}}^\natural$
is also tachyonic
(with vacuum energy $-1/e_3$) where $g^{k_3}$ is of order
$e_3=e_1e_2/(e_1,e_2)^2$.

\noindent Condition (I) is required to ensure that $\Tgt$ has the correct
residue and pole strength at any singular cusps
whereas condition (II) ensures that the composition of two Atkin-Lehner
involution invariances of $\Tgt$ is another
Atkin-Lehner invariance as in (A.4).

The Vacuum Properties are easily understood for $g$ of prime order $p$ as
follows. As described above,
$\Tgt$ is always  $\Gp$ invariant. The fundamental region for this group,
${\cal F}_p=H/\Gp$, has two cusp points at $\tau =\infty\ (q=0)$,
where $\Tgt$ has
 a simple pole and $\tau=0 $, at which $\Tgt$ may have a second pole
determined by the sign of the vacuum energy $E_0^g$ and residue given by $N_g$
from (4.1).
Thus $E_0^g$ is non-negative if and only if $\Tgt$ has a unique simple pole at
$q=0$ i.e. $\Tgt$ is a hauptmodul for $\Gp$ and $g=p-$.
 For $g=p+$ where  $\Tgt$ is invariant under
the Fricke involution $W_p:\tau\rightarrow
-1/p\tau$, then  $\MMtextorbsqr{g}{1}(\tau)=
\MMtextorbsqr{1}{g}(p\tau)$ and we have  $N_g=1$ and $E_0^g=-1/p$,
as given in the Vacuum Properties. Conversely, if
$N_g=1,\ E_0^g=-1/p$ then $f(\tau)=\Tgt-T_g(W_p(\tau))$ is  $\Gp$
invariant without any poles. Therefore $f(\tau)$ is holomorphic on the
compactification of ${\cal F}_p$ (a compact Riemann surface)
which is impossible unless $f$ is constant. But
$f(W_p(\tau))=-f(\tau)$ implies  $f=0$. Therefore, $\Tgt$ is $\Gp+$ invariant
and has a unique simple pole at $q=0$ on $H/\Gp+$ and thus
$\Gp+$ is a genus zero modular group with hauptmodul $\Tgt$.
A similar argument to this  applies in the more general
situation where $g$ is not of prime order and  $\Tgt$ can be
invariant under other Atkin-Lehner involutions
\Tuiteone. In addition, the Vacuum
Properties imply that Thompson series obey the power-map formula which
relates $\Gg$ to $\G_{g^k}$. This is an empirical observation in \CN\  not
derivable from the genus zero property \Tuiteone.

For the 82 Fricke classes $f=n+n,e_2,...$,  $\MMtextorbsqr{1}{f}(\tau)=
\MMtextorbsqr{f}{1}(\tau/n)=q^{-1/n}+0+O(q^{1/n})$.
Thus, despite the fact that $E_0^f=-1/n$,
$\Vf$ contains no massless operators because
 the first excited states of ${\cal H}_f$ with energy $1/n$
are created by the action of conformal weight 2 \ops of $\MM$
on the  $f$ twisted vacuum.
We may then repeat the argument of \S 3
to conclude that no massless \op $\psi^{(0)}(z)$ present in any other
twisted sector $\Vfk$ can be invariant under the  $\Pf$ projection (otherwise
$\psi^{(0)}(e^{2\pi i}z)=f\psi^{(0)}(z)f^{-1}=
\psi^{(0)}(z)$ obeys the defining monodromy condition for a massless \op
twisted by $f$ which is impossible). Hence, for these Fricke classes,
$\MMforb$ contains no massless \ops so that $Z_{\rm orb}(\tau)=J(\tau)$ again.
Therefore, given the uniqueness of $\MM$, we find that $\MMforb\cong\MM$.
We have therefore shown that
orbifolding $\MM$ with respect to the 38 non-Fricke classes $\{{a^*}\}$ gives
$\VL$ whereas orbifolding $\MM$ with respect to the 82 Fricke classes
$\{f\}$ reproduces $\MM$, assuming that $\MM$ is unique and the
Vacuum Properties hold  (i.e. the Thompson series
are hauptmoduls). Thus we have
$$
\eqalign{
 \VL \
\matrix{{\buildrel a\over \longrightarrow}  \cr
	    {\buildrel {a^*}\over \longleftarrow}}\
\MM\   {\buildrel f\over \longleftrightarrow}\ \MM
}
\eqno (4.5)
$$
where each arrow represents an orbifolding with respect to the denoted \aut.
We will refer to (4.5) as the Unique Orbifold Partner Property for $\MM$.

\subsect{4.2 Monstrous Moonshine from the Unique Orbifold Partner Property.}
We will now argue that the converse to the statement above is also true i.e.
assuming that $\MM$ is unique and (4.5) holds for all meromorphic
orbifoldings of $\MM$
with respect to $g\in M$, then the Vacuum Properties hold
and hence each Thompson series $\Tgt$ is a hauptmodul
for a genus zero modular group.

We begin with an orbifolding of $\MM$ with respect to an \aut, which we
denote by  $a^*$,  which produces the Leech theory $\VL$.  $a^*$ is dual to an
\auto $a$ of $\VL$ which must belong to one of the 38 classes
described in \S 3. However, assuming
the uniqueness of $\MM$, then there must be
exactly 38 different corresponding classes of \autos $\{{a^*}\}$ of $\MM$ with
Thompson series $T_{a^*}(\tau)=1/\eta_\a(\tau)-a_1$.
The associated twisted sector ${\cal V}_{{a^*}}^\natural$ therefore
has vacuum energy $E_0^{{a^*}}=0$ (and degeneracy
$N_{{a^*}}=-a_1$) in agreement with the Vacuum Properties concerning
${\cal V}_{{a^*}}^\natural$.  Furthermore,  $T_{{a^*}}(\tau)$ is known to be a
hauptmodul for the
genus zero modular group $\Gn+e_1,e_2,...$, $e_i\not =n$ and hence ${a^*}$ is
a non-Fricke element of type $n+e_1,e_2,...$ . Thus the remaining Vacuum
Properties concerning ${\cal V}_{{a^*}^k}^\natural$ must
also hold for these elements.
We will briefly consider further reasons for this result later on
in the light of our discussion of the Fricke elements.

Let us now consider the remaining allowed orbifoldings of $\MM$ with respect to
\auts, which we denote by $\{f\}$, which are assumed to reproduce $\MM$.
 Each orbifolding
is necessarily free of global phase anomalies and hence, as described above,
$\Tft$ is $\Gn$ invariant where $f$
is of order $n$. We will show that the Vacuum
Properties hold for these \aut s and that $\Tft$ is a hauptmodul which
is Fricke invariant.

$\MMforb\cong \MM$  implies the absence of massless \ops in $\Vf$.
Therefore the twisted vacuum energy  obeys  either
$E_0^{f}> 0$ or $E_0^{f}=-1/n$ (so that
$\Vf$ is tachyonic).  The first case is the only possibility in a regular
lattice orbifolding as in \S 3. $E_0^{f}=-1/n$ is also possible
for an orbifolding of $\MM$
because the lowest excited energy \ops  $\{\psi_2(z)\}$ of $\MM$
are of conformal dimension 2.
Based on our experience with lattice orbifoldings,
we expect the first excited states of ${\cal H}_f$ to be created by the action
of some of these \ops on the
twisted vacuum as in (4.3). These excited states can then
have minimum energy $1/n$ so that the absence of any massless \ops in $\Vf$
is directly due to a similar absence in $\MM$.
On the other hand,  any other negative value of $E_0^{f}$ would result in
massless \ops in $\Vf$. We will directly observe this situation below in \S 4.4
when we consider \autos based on lattice \autos for which $\Vf$
can be explicitly
constructed.  Thus, we have determined that for any
$f\in M$ either $E_0^{f}>0$
or $E_0^{f}=-1/n$ ($\Vf$ is tachyonic) whereas for $a^*\in M$,
 $E_0^{a^*}=0$. (Later on
we will eliminate the possibility of $E_0^{f}>0$ by studying
the singularities and modular properties of $\Tft$). As described before,
the behaviour of  $\Tft$ at a cusp
point $a/c$  is determined by $\MMtextorbsqr{f^{-d}}{f^c}$
(where $ad-bc=1$) with
singular behaviour when  $E_0^{f^c}<0$ where
$f^c$ is of order $n'$.  Therefore  ${\cal V}_{\rm orb}^{f^c}\cong \MM$
with $E_0^{f^c}=-1/n'$ and
the residue of this pole is $N_{f^c}$, the vacuum degeneracy of the
twisted sector ${\cal V}_{f^c}$.  We will next show that $N_{f^c}=1$.

As was the case for the lattice
orbifold constructions of \S2 and \S 3,
we may identify an \auto ${f^*}$, which is dual
to the \auto $f$, where the \ops of $\Vfk$ are eigenvectors with eigenvalue
$\omega^k$ for $\omega=e^{2\pi i /n}$. ${f^*}$ is then an \auto of the OPA for
$\MMforb$ where $\MMforb\cong \MM$ by assumption
 i.e. $f^*\in M$ and  ${\cal V}_{\rm orb}^{f^*} \cong \MM$.
We can then calculate the Thompson series $T_{{f^*}}(\tau)=
\Tr{{\cal H}_{\rm orb}^f}{{f^*} q^{L_0}}=
\sum_{k=1}^n \omega^k\MMtextorbsqr{\Pf}{f^k}$ which is $\Gn$ invariant
using the usual modular transformation properties of these traces. Furthermore,
we can show that $T_{{f^*}}(\tau)=\Tft$ by
considering the sum of Thompson series
$ \sum_{k=1}^n T_{f^{*k}} (\tau)=
\sum_{k=1}^n \Tr{{\cal H}_{\rm orb}^f}{{f^*}^k q^{L_0}}$. Since only the
untwisted sectors contribute we find
$$
\eqalign{
\sum_{r\vert n} d_r  T_{f^{*r}}(\tau) =\sum_{r\vert n} d_r  T_{f^{r}}(\tau)
}
\eqno(4.6)
$$
where $d_r$ is the number of integers $k\in \{1,...,n\}$ with $(k,n)=r$ so that
$T_{f^r}(\tau)=T_{f^{k}}(\tau)$ and likewise for $f^*$. For $n=p$, prime,
we have $d_1=p-1,\  d_p=1$ and (4.6) implies that $T_{{f^*}}(\tau)=\Tft$.
For $n$ not prime we may identify the singularities
of $\Tft $ and $T_{{f^*}}(\tau)$
as follows.  Consider the modular function $\phi(\tau)=d_1(T_{f^*}-T_f)$.
As described above, the behaviour of $\phi(\tau)$ at $\tau=0$
can only be singular if either $E_0^{f^*}=-1/n$ or
$E_0^{f}=-1/n$ or both where $\phi(-1/\tau)=Aq^{-1/n}+0+...$ for $A=N_{f^*}$ or
$N_f$ or $N_{f^*}-N_{f}$ respectively.
But from (4.6), $\phi(\tau) =\sum_{r>1} d_r  (T_{f^{r}}-T_{f^{*r}} )$
has  singular behaviour at $\tau=0$ determined by
 $\phi(-1/\tau)=Bq^{-r/n}+...$ which is
inconsistent unless $E_0^{f^*}=E_0^{f}=-1/n$ and $N_{f^*}=N_{f}$ for all
tachyonic sectors.  Therefore $\phi(\tau)$ is $\Gn$
invariant without singularities
and defines a holomorphic function on the compactification of $H/\Gn$
(a compact Riemann surface). This
is impossible unless $\phi(\tau)$ is a constant
which must be zero since Thompson series contain no constant term.
Therefore $T_{{f^*}}(\tau)=\Tft$ and so $f$ and $f^*$
can be identified as members of the same conjugacy class of $M$ (apart from the
classes $27A,\ 27B$ where possibly $f$ and $f^*$ are in different classes).

We next examine the centraliser $C({f^*}\vert M)$ by a similar analysis to
that of \S 3 and Appendix B.  Define ${Aut}(\Pf\Vf)$ to be
the \auto group of the
OPA for $\MMforb$ which maps $\Pf\Vf$ into itself. Then
$n.{Aut}(\Pf\Vf)\subseteq C({f^*}\vert M)$ where the extension is the
central cyclic group  generated by ${f^*}$.
{}From (4.3), the vacuum \ops $\{\sigma_{f}^r(z)\}$ of $\Vf$ must form a
$N_{f}$
dimensional representation for
$C(f\vert M)/\langle f\rangle $ which  defines some extension $L_\sigma$
so that ${Aut}(\Pf\Vf)=L_\sigma.(C(f\vert M)/\langle f\rangle )$.
Therefore we find that $n.L_\sigma.(C(f\vert M)/\langle f\rangle)
\subseteq C({f^*}\vert M)$.
However, this is impossible since ${f^*}$ and $f$ are in the same conjugacy
class of $M$ unless $L_\sigma=1$ so that the twisted vacuum
of $\Vf$ is unique where $N_{f}=1$. (For the two classes $27A,\ 27B$, the
centralisers are of the same order so that again $L_\sigma=1$).

We have  shown that for any $f\in M$ where $\MMforb\cong \MM$,
$\Vf$ has vacuum energy $E_0^f>0$ or $E_0^f=-1/n$ with degeneracy $N_f=1$.
We will now eliminate the possibility of $E_0^{f}>0$.
If $E_0^{f^k}\ge 0$ for all $k\not =n$,
then $\Tft$ has a unique simple pole at $q=0$ and is therefore a
hauptmodul for $\Gn$. This is only possible for
 $2\le n \le 10,\ n= 12,13,16,18$ with
hauptmodul $\Tft=1/\eta_\a(\tau)-a_1$ for the corresponding \auto
$\a\in {\rm Co}_0$ in Table 1 with modular group $\Gn=n-$.
Then under $S:\tau\rightarrow -1/\tau$ we get
$E_0^f=0$ with $N_f=-a_1\not =0$ in contradiction so that $E_0^f=-1/n$
in these cases which includes all the prime ones.
For the remaining non-prime cases with
some $E_0^{g^k}<0$, we consider the composition of two
orbifoldings of $\MM$ which will allow us to
determine the location and strength
of any  singularities of $\Tft$.

Choose $f\in M$ of non-prime order $n$ (where either $E_0^f>0$ or $E_0^f=-1/n$)
such that for any $f_1\in M$ of order $n_1<n$ where
${\cal V}^{f_1}_{\rm orb}\cong\MM$ then $E_0^{f_1}=-1/{n_1}$.
This choice includes the \auto $f$ of least order with $E_0^f>0$ which
we will  show cannot exist.  With this choice of $f$,
if $\MMfrorb\cong\MM$ for $f^r$ of order $e=n/r$,
$\Vfr$ must be  tachyonic with  $E_0^{f^r}=-1/e$.
(We may assume that $r\vert n$
since $\Vfr$ and ${\cal V}_{f^{r'}}^\natural$ are isomorphic for $(n,r)=(n,r')$
in general).
We will show that $\Vfe$ must also be tachyonic with $E_0^{f^e}=-1/r$ where
$(e,r)=1$ i.e. $e\vert\vert n$. This
corresponds to the singularities given in (I) of the Vacuum Properties and will
also lead to the closure property in (II) once we have shown that $E_0^f=-1/n$.
In constructing $\MMforb\cong\MM$ we employ twisted \ops which are
also involved in constructing $\MMfrorb\cong\MM$. The contribution to
$\MMforb$ from these \ops is
$$
\eqalign{
\Pf(\MM\oplus\Vfr\oplus...\oplus{\cal V}_{f^{r(e-1)}}^\natural)=
{1\over r}(1+f+ ...+f^{r-1})\MMfrorb\cong {\cal P}_{f' }\MM
}
\eqno (4.7)
$$
where $f'$ is an \auto of $\MM$ of order $r$
defined by the \auto $f$ acting on $\MMfrorb$ (since $f^r$ acts as unity on
$\MMfrorb$).  But ${\cal P}_{f' }\MM$ is the untwisted
contribution to the orbifolding of $\MMfrorb$ with respect to
$f'$. Furthermore, the orbifolding of $\MM$ with respect to $f$ is
a composition of the
orbifolding of $\MM$ with respect to $f^r$ and the orbifolding of
$\MMfrorb\cong\MM$ with respect to $f'$ as follows
$$
\eqalign{
\matrix{ \MM  \cr
{\buildrel f^r\ \  \over\nearrow} \qquad {\buildrel \ \ f'\over\searrow}  \cr
 \MM\quad  {\buildrel f\over \longrightarrow} \quad\MM\cr}
}
\eqno (4.8)
$$
where the arrows represent an orbifolding with repect to the denoted
\aut.  Thus ${\cal V}_{\rm orb}^{f'}\cong \MM$ and therefore
$\Vfhat$ is also tachyonic with $E_0^{f'}=-1/r$ by our
choice of $f$ since $f'$ is of order $r<n$. We can check for the
consistency of this composition of orbifoldings by considering the Thompson
series $T_{f'}(\tau)$ for $f'$ as a trace over $\MMfrorb$.
Under $S:\tau\rightarrow -1/\tau$ this becomes
$$
\eqalign{
T_{f'}(-{1\over\tau})=\sum_{k=1}^e \MMorbsqr{\Pfr}{f^{1+rk}}
}
\eqno(4.9)
$$
which must have leading behaviour $q^{-1/r}+...$ from (4.1).
Therefore, at least one
of the twisted sectors contributing to the RHS of (4.9) must be tachyonic with
vacuum energy $-1/r$  and $f^{1+rk}$  of order $r<n$. Thus
$r(1+rk)=nl$ for some $l$ so that $el-rk=1$ which
implies that $(e,r)=1$. Therefore,
$e\para n$ (and $r\para n$) and ${\cal V}_{f^{e}}^\natural$ is tachyonic with
vacuum energy $-1/r$ (as is
the isomorphic twisted sector ${\cal V}_{f^{el}}^\natural$ since $(l,r)=1$).
Thus orbifolding
$\MM$ with respect to $f^e$ also reproduces $\MM$. To summarise,
for $f$ of order $n$ as chosen, if $\MMfrorb\cong\MM$
(so that  $\Vfr$ is tachyonic)
where  $f^r$ is of order $e=n/r$  then
$e\para n$ and ${\cal V}_{f^{e}}^\natural$ must also be tachyonic with
${\cal V}_{\rm orb}^{f^e}\cong \MM$.

This translates into information about the singularity structure of $\Tft$
 \Tuiteone.
If we choose the representative form for the Atkin-Lehner (AL)
involution  $W_e=(\matrix{e & b\cr
			  n &  de\cr})$, for $e\not = n$, as in Appendix A. Then
$T_f(W_e(\tau))=\MMtextorbsqr{f^{-ed}}{f^r}(e\tau)=q^{-1}+0+O(q)$
 when $\MMfrorb\cong\MM$. Note that the constant term is zero
since $\Vfr$ contains no massless operators.
We define $\tau_e=W_e(\infty)=1/r$ which
we call an AL cusp.  On the fundamental region  ${\cal F}_n=H/\Gn$,
the singularity at $\tau_e$ is then a simple pole since
$W_e$ is an \auto of ${\cal F}_n$.
In addition,  $\Tft$ also has a simple pole at
the AL cusp $\tau_r=1/e=W_r(\infty)$ since ${\cal V}_{\rm orb}^{f^e}\cong \MM$.
Thus  $\Tft$ has simple poles with residue 1
at $\tau =\infty \ (q=0)$ and
possibly at $\tau=0$ (if $E_0^f=-1/n$) and
at the AL cusps $\tau_e$ and $\tau_r$.

We next show that $\Tft$ must always be singular at $\tau=0$ with $E_0^f=-1/n$.
Suppose that $E_0^f>0$, then under the Fricke
involution $W_n:\tau\rightarrow -1/n\tau$, $\tau_e$ and $\tau_r$
 are interchanged. Then $\phi(\tau)=\Tft-T_f(W_n(\tau))$ is a $\Gn$ invariant
meromorphic function on ${\cal F}_n$ with two simple poles at
$\tau=\infty $ ($q=0$)
and $\tau=0$. $\phi(\tau)$ also has zeros at $\tau_e$ and $\tau_r$ since
$\phi(W_e(\tau))=q^{-1}-q^{-1}+0+O(q)$ where it is
essential that the  AL poles  have the same strength and residue and
that $\Vfr$ and ${\cal V}_{f^e}^\natural$ contain no massless operators.
Likewise, $\phi(\tau)$ has zeros at any other such pairs of singular AL cusps.
 But $\phi(\tau)$
is odd under $W_n$ and therefore also has a zero at the $W_n$ fixed point
$\imath/\sqrt n$. Thus $\phi$ has two simple poles  and at least
three zeros on the
compactification of ${\cal F}_n$ which is a compact Riemann surface. But every
meromorphic function on a compact Riemann surface has an equal number of zeros
as poles. Therefore, there is a contradiction and hence $E_0^f=-1/n$.

We have now derived  condition (I) of the Vacuum Properties for $f$.
In addition, a restricted version
of the AL closure condition (II) has also been demonstrated. Namely,
if $\Vf$ and
${\cal V}_{f^r}^\natural$  are tachyonic (where $f^r$ is of order $e\para n$),
then so is
${\cal V}_{f^e}^\natural$ where $f^e$ is of order $r=ne/e^2$.
We can use this to generate
the general AL closure property as follows.
Suppose that $\Vfone$ and $\Vftwo$ are both
tachyonic with $f_1=f^{r_1}$ and $f_2=f^{r_2}$ with $r_1\not = n/r_2$
where $f_i$ is of order $e_i$ where  $e_i\para n$.
(We can take $r_i\vert n$, as before, since
$\Vfr$ and ${\cal V}_{f^{r'}}^\natural$ are isomorphic for $(n,r)=(n,r')$).
Then the  sectors twisted by $f^{e_i}$ of order $r_i$ are also tachyonic.
 By interchanging $e_i$ with $r_i$ if necessary we
can assume that $(e_1,e_2)=1$.
This is easily shown by observing
that the order $n$ of every element of $M$ has at
most 3 distinct prime divisors ($n<2.3.5.7=210$).
Then $e_3\vert\vert n$ for $e_3=e_1 e_2$ and $r_3=r_1r_2/(r_1,r_2)^2=n/e_3$
with $r_1=r_3 e_2$ and $r_2=r_3 e_1$.
Consider $g=f^{r_3}$ of order $e_3$.
Then $g^{e_1}=f^{r_2}$  and
$g^{e_2}=f^{r_1}$ are of order $e_2$ and $e_1$, respectively,
so that the corresponding twisted sectors are tachyonic.  Therefore,
 by taking the  composition of orbifoldings
with respect to $g^{e_i}$, as in (4.8),
we find that $\Vg$ is also tachyonic with
$g=f^{r_3}$ of order $e_3=e_1 e_2$.  As before, the  sector twisted
by $f^{e_3}$ of order $r_3=r_1 r_2/(r_1,r_2)^2$ must also then be tachyonic.
Thus the general AL closure condition (II) is derived.

We have now demonstrated that the genus zero property for Thompson
series can be derived from (4.5) assuming that $\MM$ is unique  and
so we have :

\proclaim  Monstrous Moonshine is equivalent to the
Unique Orbifold Partner Property.
Assume that the FLM uniqueness conjecture holds. Then  $T_g(\tau)$ for
$g\in M$ is a hauptmodul for a genus zero modular group $\Gn+e_1,e_2,...$ if
and only if the only meromorphic orbifoldings of $\MM$ with
respect to $g$ are $\VL$ and $\MM$.

We note that we may also understand the Vacuum Properties already
found for the non-Fricke elements $a^*$ dual to $a$ in a similar fashion
to this derivation for the Fricke elements. Suppose that
$f=a^{*r}$ of order $e=n/r$ is Fricke so that
$\MMforb\cong\MM$. We can then
deduce that $e\para n$ and that $a^{*e}$ is non-Fricke as follows.
The  orbifolding of $\MM$ with respect to $a^*$ (which gives
$\VL$) is the
composition of the orbifolding of $\MM$ with respect to $f$ and
the orbifolding of $\MMforb\cong\MM$ with respect to $b^*$ of order
$r$ where $b^*$ is the action of $a^*$ on $\MMfrorb$. Thus $b^*$ is
dual to $b$, one of the 38 \autos of $\VL$ discussed in \S 3.
It is straightforward to then see that $b=a^e$ (lifted from $\a^e$)
has the correct
action on $\MMfrorb$  to be dual to $b^*$. If we examine the
38 \autos listed in Table 1, we find that $\a^e$ is contained
in Table 1 if and only if $e\para n$ and $\eta_\a(\tau)$ is
invariant under the AL involution $W_e$
(but is inverted by $W_r$). In fact, in each such case this
follows from the symmetry properties of the characteristic
equation parameters where $ a_k=-a_{n/k}=a_{ek_r/k_e}=-a_{rk_e/k_r}$ (with
$k_e=(k,e)$ and $k_r=(k,r)$) so that $\b=\a^e$ has parameters
$b_k=-b_{r/k}$. Similarly, the closure condition (II) follows
directly from these parameter relationships.

\subsect {4.3 Moonshine for $\bf n\vert h+e_1,e_2,...$, $\bf h\not=1$.}
Let us now consider the Thompson series for the classes of $M$ which cannot
be employed to construct a meromorphic  modular invariant orbifold
 due to a global phase  anomaly.
These classes consist of the 13 non-Fricke classes
of \S 3 and 38 Fricke classes. The twisted
sector $\Vg$ for the non-Fricke classes and some of the Fricke classes can be
constructed since they belong to the centraliser $C(i\vert M)=2^{1+24}.\Co$
as described below in \S 4.4. We find that $E_0^g=1/nh$ for the non-Fricke
classes and $E_0^g=-1/nh$ for the Fricke classes where $h\vert n$.
We will assume that this latter property is also
correct for the remaining Fricke
classes. The integer $h\not =1$ parameterises the global phase anomaly present
in these cases where $T^n:\MMtextorbsqr{1}{g}\rightarrow e^{\pm 2\pi i/h}
\MMtextorbsqr{1}{g}$. In \S 3 we considered the 13
Leech lattice \autos with a global phase anomaly where we found an isomorphism
between $\Ha\otimes...\otimes\Ha$ and $\Hah$ in (3.13). A similar isomorphism
is also expected here between the twisted Hilbert spaces
$\Hg$ and $\Hgh$ as follows
\Tuiteone. Let $\tilde\psi_i(z) $ create a twisted state in $\Hg$
by acting on the twisted vacuum states
$\{\vert \sigma_g^l\rangle \}$. Then $\tilde \Psi_{i_1...i_h}(z)=
\tilde\psi_{i_1}(z^h)\otimes...\otimes\tilde\psi_{i_h}(z^h)$
which acts on $\vert \Sigma_{g^h}^{l_1...l_h}\rangle =
\vert \sigma_g^{l_1}\rangle \otimes...\otimes
\vert \sigma_g^{l_h}\rangle $ obeys the monodromy
condition (4.2) for $g^h$ of order $n/h$. $\tilde \Psi(z)$
creates a state in $\Hgh$ but is not a primary conformal field.
 For the non-Fricke classes the states $\vert \Sigma_{g^h}\rangle $
 are of energy $h/n$ whereas for the Fricke classes,
$\vert \Sigma_{g^h}\rangle $ is unique and is of energy $-h/n$
 and reproduces the vacuum of $\Hgh$. Thus as before, the
global phase anomaly disappears by taking such a tensor product.
Thus an identification can be made between the non-massless
states of $\Hgh$ and $\Hg\otimes...\otimes\Hg$.
For the non-Fricke classes, $\Hgh$ always contains $N_{g^h}>0$
massless states whereas
the energies of all the states of $\Hg\otimes...\otimes\Hg$ are positive.
On the other hand, for the Fricke classes, $\Hgh$  contains no
massless states but
$\Hg\otimes...\otimes\Hg$ contains $hN_1$ massless states
where $N_1$  is the number of \ops of $\Hg$ with
first excited energy level $-1/nh+1/n$.
Therefore the partition functions are expected to be related as follows:
$$
\eqalign{
[\MMorbsqr{1}{g}(h\tau)]^h=\MMorbsqr{1}{g^h}(\tau)+C
}
\eqno (4.10)
$$
where $C=-N_{g^h}$ for the non-Fricke classes and $C=hN_1$ for the Fricke
classes. In terms of the Thompson series this is the harmonic
formula of Conway and Norton \CN
$$
\eqalign{
[T_g(\tau/h)]^h=T_{g^h}(\tau)+C
}
\eqno(4.11)
$$
This relationship implies that $\Tgt$ is
$\Gnhe$ invariant up to $h$ roots of unity.
We also know that $\MMtextorbsqr{1}{g}(\tau)$ is $T^{nh}$ invariant from which
we may show that $\Tgt$ is $\GN$ invariant with $N=nh$.
Thus $\Gnh$ must be in the normaliser of $\GN$ and hence $h\vert 24$
from Appendix A. The invariance group $\Gg$ for $\Tgt$ of index $h$ in $\Gnhe$
can then be shown to be of genus zero with hauptmodul $\Tgt$ because the
invariance group $\G_0({n\over h})+e_1,e_2,...$ of $T_{g^h}(\tau)$ is of
genus zero \Tuiteone.

\subsect{4.4 Twisted operators for $\bf c\in C(i\vert M)$.}
We will now discuss the construction of
the twisted sector $\Vc$ for $c\in C(i\vert M)$ where $c$  is lifted
from a Leech lattice \auto $\c\in {\rm Co}_0$ and
is therefore geometrical in origin.  Because $c$ does not interchange
 the sectors $\Pr \VL$ and $\Pr \Vr$ in the original FLM construction,
the Thompson series for $c$
can be explicitly computed \refs{\FLMzero,\FLM,\FLMb} to be
$$
\eqalign{
T_c(\tau)=&\orbsqr{c\Pr}{1}+\orbsqr{c\Pr}{r}\cr
=&{1\over 2}\lbrace{\Theta_{\L_\c}(\tau)\over \eta_\c(\tau)}+
{\Theta_{\L_{-\c}}(\tau)\over \eta_{-\c}(\tau)}
+\Tr{}{c_T}{\eta_\c(\tau)\over \eta_\c(\tau/2)}
-\Tr{}{c_T}{\eta_{-\c}(\tau)\over \eta_{-\c}(\tau/2)}\rbrace\cr}
\eqno(4.12)
$$
where $\Theta_{\L_{\pm\c}}$ is the theta function
 for the sublattice $\L_{\pm\c}$ of $\L$ invariant under $\pm\c$
and $\eta_{\pm \c}$ is the eta function as in
(3.5a). The lifting
of $\c$ to an \auto $c$ of $\VL$ is
chosen so that $c c(\beta) c^{-1}=c(\beta)$ for
all $\beta\in \L_\c$ (see (3.1a)) and similarly
for $rc$ lifted from $-\c$ (where
$r$ and $c$ commute).
$c_T$ is the action of the lifting of $\c$ on the vacuum of $\Vr$.
Given the usual modular transformation properties for the traces of (4.12),
$T_c(\tau)$ is automatically $\G_0(m)$ invariant (up to possible phases)
where $m$ is the order of $\pm \c$ in $\Co$.
We also find that under $S:\tau\rightarrow -1/\tau$
$$
\eqalign{
T_c(-1/\tau)=& {1\over 2}\lbrace \orbsqr{1}{c}+\orbsqr{r}{c}
+\orbsqr{1}{rc}+\orbsqr{r}{rc}\rbrace\cr
=&{1\over 2}\lbrace
{D_\c^{1/2}\over V_\c} {\Theta_{\L_\c^*}(\tau)\over \eta^*_\c(\tau)}+
{\Tr{}{c_T}\over 2^{d_{+}/2}}{\eta^*_\c(\tau)\over \eta^*_\c(2\tau)}\cr
&+{D_{-\c}^{1/2}\over V_{-\c}}
{\Theta_{\L_{-\c}^*}(\tau)\over \eta^*_{-\c}(\tau)}
-{\Tr{}{c_T}\over 2^{d_{-}/2}}
{\eta^*_{-\c}(\tau)\over \eta^*_{-\c}(2\tau)}\rbrace
}
\eqno(4.13)
$$
where $\Theta_{\L_{\pm \c}^*}(\tau),\ \eta^*_{\pm \c}(\tau)$ and $D_{\pm \c}$
are defined  as (3.5) and $V_{\pm \c}$ is the volume of
$\L_{\pm \c}$. $d_{\pm}=\sum_k c^{\pm}_k$
determines the number of unit eigenvalues of
${\pm}\c$ with characteristic equation
parameters $\{ c_k^{\pm}\}$ as in (3.2).  From (4.1), we may
 therefore define the twisted sector for each such $c\in C(i\vert M)$ to be
$$
\eqalign{
\Vc=\Pr\VLc\oplus\Pr\VLrc
}
\eqno(4.14)
$$
where $\VLc$ and $\VLrc$ are the twisted sectors constructed in the standard
way from the $\L$ compactified string as described in \S 3.4 and Appendix B
\refs{\Lepowsky,\KacP,\Dixon,\CHZn} where
$\tilde X(e^{2\pi i}z)=\pm(\c)^{-1}\tilde X(e^{2 \pi i}z)+2\pi \beta$.
Then $\chi_c \in \VLc$ and $\chi_{rc}\in \VLrc$ obey
the monodromy conditions $\chi_c(e^{2\pi i }z)=c\chi_c(z)c^{-1}$ and
$\chi_{rc}(e^{2\pi i }z)=rc\chi_{rc}(z)(rc)^{-1}$ as in (3.12). For
$\psi_r\in \Vr$ we  expect the (schematic) OPAs
$\psi_r\chi_{c}\sim\chi_{rc}$ and  $\psi_r\chi_{rc}\sim\chi_{c}$
to hold together
with the usual OPAs of (2.5), (2.15) and (3.14). Since $r$ and $c$ commute, $r$
preserves these OPAs and hence the projection with respect to $\Pr$ can be
taken. Then for $\psi\in \MM= \Pr(\VL\oplus\Vr)$, the monodromy
conditions and OPA (4.2) and (4.3) follow
where $\{ \sigma_c\}$ denotes the vacuum \ops for $\Pr(\VLc\oplus\VLrc)$.
Thus $\Vc$ given in (4.14) satisfies  the defining relations for the $c$
twisted sector.

We may check for the other properties satisfied by $\Vc$
(particularly when $c$ is a Fricke element of $M$)
which lead to Thompson series
which are hauptmoduls as described in \S4.2 and \S4.3. In \Lang\ a
survey is presented of the modular functions $\textorbsqr{c}{1}=
\Theta_{\L_\c}/\eta_\c=q^{-1}+c_1+...$ for all $\c \in \rm Co_0$.
It is shown that
$\Theta_{\L_\c}/\eta_\c$ is a  hauptmodul for a genus zero fixing group
$n\vert h+e_1,e_2..$ for all but
15 classes of $\rm Co_0$ (thereby falsifying a conjecture of
Conway and Norton \CN).
We will return to these anomalous classes
below.  For the remaining classes,  we may describe some general properties
of the vacuum of $\VLc$,  similar to the vacuum properties of $\Vg$ above.
Thus $\Theta_{\L_\c}/\eta_\c$ is Fricke
invariant under $\tau\rightarrow-1/nh\tau$
if and only if the vacuum energy of $\VLc$ obeys $E_0^c=-1/nh$
 and the vacuum degeneracy $N_c=D_\c^{1/2}/V_\c=1$. (We will
call the corresponding class of $\rm Co_0$ a Fricke class). Otherwise,
$E_0^c\ge 0$ and the vacuum may be degenerate.
Likewise, the other vacuum properties of \S 4.1 must hold.

For all the Fricke classes, the characteristic equation
parameters $c_k$ are observed to
obey the symmetry condition $c_k=c_{nh/k}$
where $h\vert k$ for all $c_k\not =0$ \Lang.
Therefore  $D_\c=(nh)^{d}$, $\eta^*_\c(\tau)=\eta_\c(\tau/nh)$ and
hence $E_0^c=-1/nh$.  Similarly, from (4.13) we find that since $N_c=1$,
$V_\c=(nh)^{d/2}$ and $\Theta_{\L_\c^*}(\tau)= \Theta_{\L_\c}(\tau/nh)$
where $\beta^2\in 2hZ,\ \beta^2\ge 4 $ for $\beta\in\L_\c\subset\L$.
Thus for $h=1$,  $\beta^{*2}\ge 4/n$  whereas for $h\not= 1$,
$\beta^{*2}\ge 2/n$ for all $\beta^*\in \L_\c^*$.
Furthermore we can observe from \KondoTasaka\  that
$\L_\c\equiv \sqrt {nh}\L_\c*$
in many such cases (e.g. for $\c=1^4 8^4/2^2 4^2$
of order $n=8$,  $\L_\c=\sqrt{2} D_4$ and
$\L_\c^*= D_4^*/\sqrt{2}\equiv D_4/2$, after a $\pi/4$ rotation).
This non-trivial property for $\L_\c$ is
very likely to be true for all such Fricke \autos of the Conway group.

{}From (4.13), the uniqueness of the $c$ twisted vacuum $\vert\sigma_c\rangle$
for the Fricke classes  implies that
$\Tr{}{c_T}=\epsilon_r 2^{d/2}$ where
$r\vert\sigma_c\rangle=\epsilon_r\vert\sigma_c\rangle$ with
$\epsilon_r=\pm 1$. For $h=1$, when $E_0^c=-1/n$, the first excited (massless)
states of this sector are given by
$\vert \psi^i_c\rangle =\tilde\alpha^i_{-1/n}\vert \sigma_c\rangle=
\lim_{z\rightarrow 0}z^{1-1/n}\partial_z \tilde X^i(z)\vert \sigma_c\rangle$
for $i=1,...,c_n$ where $\partial_z \tilde X^i(z)$ is an $\omega^{-1}$
eigenvector of $\c$ which implies
 $r\vert \psi^i_c\rangle=-\epsilon_r\vert \psi^i_c\rangle$.
Since $\beta^{*2}\ge 4/n$, no massless states
are associated with  the dual lattice $\L_\c^*$.
Hence, for any Fricke class $\c\in\rm Co_0$ with $h=1$, we have
either $\textorbsqr{\Pr}{c}=q^{-1/n}+0+O(q^{1/n})$ for $\epsilon_r=1$ or
$\textorbsqr{\Pr}{c}=c_1+O(q^{1/n})$ for $\epsilon_r=-1$.
For the Fricke classes with $h\not =1$, the first excited states of $\VLc$
with energy
$-1/nh+1/n$ are given by $\vert \psi^i_c\rangle$ above together with
states $\vert\beta^*\rangle$ created by $e^{i \langle\beta,\tilde X(0)\rangle}$
for  $\beta^{*2}=2/n$.
Thus for $h=1$, $\Pr\VLc$ contains
either a unique vacuum with energy $E_0^c=-1/n$
but no massless \ops ($ \epsilon_r=1$) or else has a massless vacuum
($ \epsilon_r=-1$). Similarly,  for $h\not =1$,  $\Pr\VLc$ contains
either a unique vacuum with  $E_0^c=-1/nh$ with
first excited  \ops of energy $-1/nh+1/n$  or else has a vacuum of energy
$-1/nh+1/n$.

We may use these observations to describe the corresponding properties of
$\Vc$ defined in (4.14).  Consider $\c$  any Fricke element of $\rm Co_0$ of
order $n$ with $h=1$.  If $n$ is odd then $-\c$ is
of order $2n$ and $\VLrc$ has vacuum
energy $E_0^{rc}=1/2n>0$. If $n$ is even then $-\c$ is of order $n$ or $n/2$
and we can observe from \Lang\ that $E_0^{rc}\ge 0$ in all cases.
For $-\c$ of order $n$ with
$E_0^{rc}=0$, one can check from (4.13) and \Lang\
that  $r \vert \sigma_{rc}\rangle
=- \epsilon_r\vert \sigma_{rc}\rangle$, with $ \epsilon_r$ as above, so that
$\Pr\VLrc$ contains no massless \ops for $ \epsilon_r=1$. If $-\c$ is
of order $n/2$ then $\c^{n/2}=\r$ and $r=c^{n/2}$ so that $ \epsilon_r=-1$
from (4.13) (by considering invariance under $\tau\rightarrow\tau+n/2$).
Thus, for
any Fricke element $\c\in\rm Co_0$ with $h=1$,
$\Vc$ contains either a unique vacuum of energy $-1/n$ and
no massless \ops so that $c\in M$ is Fricke or
$\Vc$ contains a massless vacuum and $c\in M$ is non-Fricke.
One can similarly show
for a Fricke class $\c\in\rm Co_0$ with $h\not =1$ that $\Vc$ either
contains a unique vacuum with energy $-1/nh$ and first excited \ops with
energy $-1/nh+1/n$  ($c$ is  Fricke in $M$) or else has a vacuum of energy
$-1/nh+1/n$ ($c$ is non-Fricke in $M$). Likewise, if $\c$ and
$-\c$ are both non-Fricke then $c$ is non-Fricke in $M$ and $\Vc$ has the
required properties.
Thus $\Vc$ defined in (4.14) possesses all the properties
for a Monster group twisted sector as described in \S 4.2 and \S 4.3.

Let us now discuss the 15 anomalous  \autos $\{\c\}$ mentioned earlier
for which
$\textorbsqr{c}{1}=\Theta_{\L_\c}/\eta_\c$ is not a hauptmodul but is
fixed by a genus zero modular group \Lang.  These classes fall into 5 families
of the form
 $\{\c_1,\c_2,\c_3\}$ with each $\c_i$ of the
same order $n=6,10,12,18$ or 30.  For each such
$\c$,  part (I) of the vacuum properties \S 4.1 is satisfied but the
Atkin-Lehner closure condition (II) fails and so $\Theta_\c/\eta_\c$
 is not a hauptmodul. For example,  for $n=6$,  $\{\c_1,\c_2,\c_3\}$
have Frame shapes
$\{1^4 2.6^5/3^4,\ 2^5 3^4 6/1^4,\ 1^5 3.6^4/2^4\}$ (where $\c_1=-\c_2$).
Then $\Theta_{\c_1}/\eta_{\c_1}$  has simple  poles with residue 1
at the cusps $\tau=\infty,\ 0$ and the AL cusp $\tau_2=1/3$ but
not at the AL cusp
$\tau_3=1/2$. Likewise,  for $\c_2$ and $\c_3$, the  poles  occur
at $\{\infty,1/2,1/3\}$ and $\{\infty,0,1/2\}$.  The other anomalous
 families have very similar properties \Lang. Despite this behaviour, one can
repeat the analysis above to show that $\Vc$ of (4.14) possesses all the
required properties given in \S 4.2.

We will end this section with some remarks concerning the reorbifolding of
$\MM$ with respect to Fricke  elements of $M$. For a Fricke element
 $c\in C(i\vert M)$ of order $m$ with $\Vc$ as in (4.14), then given (4.5),
we find  $\MM={\cal V}_{\rm orb}^c=
{\cal P}_c(\MM\oplus\Vc\oplus...\oplus {\cal V}_{c^{m-1}}^\natural)$ is just a
$Z_2\times Z_m$ orbifolding of $\VL$ with respect to the abelian group
generated
by $r$ and $c$.  We can similarly expect that the observations
of this subsection
can be generalised  to the other assumed constructions of $\MM$ given in \S 3
based on the 38 \autos $\a$ of Table 1.
Thus for $c\in C(a^*\vert M)$, we can define
$\Vc=\Pa\VLc\oplus...\oplus\Pa{\cal V}_{ca^{n-1}}$, where $a$ and $c$ commute,
which satisfies the monodromy
conditions and OPA of
(4.2) and (4.3). Then reorbifolding ${\cal V}_{\rm orb}^c$
with respect to an element of $C(a^*\vert M)$
is equivalent to a $Z_n\times Z_m$ orbifolding of $\VL$ with respect
to the abelian group generated by $a$ and $c$. Thus,  assuming (4.5) so that
${\cal V}_{\rm orb}^c=\MM$ for a Fricke element $c\in C(a^*\vert M)$,
we can, in principle, provide a large family of $Z_n\times Z_m$
orbifold constructions of $\MM$ from $\VL$.

\beginsection{5.  Concluding remarks.}
We  conclude with a number of observations concerning various open
questions and some generalisations of the constructions considered
above.
We begin with a few remarks about Norton's Generalised Moonshine
\Mason\ which concerns Moonshine for modular functions  associated with
 centraliser groups of elements in the Monster. In \DGH\ it was
suggested that these correspond to orbifold traces of the form
$\MMtextorbsqr {g_1}{g_2}$ for $g_1\in C(g_2\vert M)$.
Given the usual modular transformation properties for such traces,
 then the structure of the vacuum of the Monster twisted sectors
${\cal V}_{g_2}^\natural$ described here should be sufficient to
to show that each such trace is a hauptmodul.
 A general discussion of this will appear elsewhere \Tuitegen\ but
we  make three brief observations here. Firstly,  for $g_2$ a non-Fricke
element, the vacuum of ${\cal V}_{g_2}^\natural$
is degenerate in most cases so that each $g_1$ is actually an element
of an extension of $C(g_2\vert M)$ in these cases,
 as observed by Norton \Norton.
For the remaining non-Fricke and all the Fricke classes,
no such extension of the centraliser is required.
Secondly,  $\MMtextorbsqr {g_1}{g_2}$
can be easily shown to be a hauptmodul for $g_1$ and $g_2$ of relatively
prime order $n_1$ and $n_2$ with
associated modular groups $n_1+e_1,e_2,...$ and  $n_2+e'_1,e'_2,...$ i.e.
$h_1=h_2=1$ where the corresponding twisted sectors are global phase
anomaly free.  Since $(n_1,n_2)=1$ we have
$n_1b+n_2a=1$ for some $a,b$. Define $g=g_1^a g_2^b$ of
order $n=n_1n_2$ so that $g_1=g^{n_2}$ and $g_2=g^{n_1}$. Then under
a modular transformation with respect to
 $\gamma=(\matrix{a  & b\cr
                              -n_1 & n_2\cr})$
we find
$$
\eqalign{
\Tgt=\MMorbsqr{g}{1}\longrightarrow
\MMorbsqr{g^{n_2}}{g^{n_1}}=\MMorbsqr{g_1}{g_2}
}
\eqno (5.1)
$$
Therefore $\MMtextorbsqr{g_1}{g_2}$ is a hauptmodul
for $\Gg=n+n,e_1,e'_1,e_2,e'_2,...$
 if $e_i=n_1$ and  $e'_j=n_2$ for some $i,\ j$ and
$\Gg=n+e_1,e'_1,e_2,e'_2,...$
otherwise i.e. $g$ is Fricke if and only if
both $g_1$ and $g_2$ are Fricke. This
property is observed for all the appropriate modular functions associated
with the centralisers of the limited number of
 elements of $M$  discussed in \Queen.
Our last observation concerns Moonshine for  $C(g_2\vert M)$ where
$T_{g_2}(\tau)$  has modular invariance group $n_2\vert h+e_1,e_2,...$
with $h\not =1$. From \S4.3 we expect that the following  harmonic formula
should hold for each $g_1$
$$
\eqalign{
\bigl\lbrack\MMorbsqr{g_1}{g_2}(h\tau)\bigr\rbrack^h=
\MMorbsqr{g_1}{g_2^h}(\tau)+C}
\eqno(5.2)
$$
where $C$ is a constant. For the case $g_2=3\vert 3$, this formula can be
verified \Queen.

The use of non-meromorphic OPAs has been central in our discussion.
Such algebras were employed both in defining the properties of twisted
\ops  and in considering reorbifoldings.  From this point of view, the two
meromorphic CFTs which are orbifold partners are embedded
in a larger set of \ops $\Vp$ obeying a non-meromorphic OPA.
 However, a rigorous construction of such a
non-meromorphic OPA has yet to be given even in the simplest $Z_2$ case.
Another interesting  question is to ask what form does  the \auto group
for $\Vp$ take ?
This  has not even been determined in the original
FLM $Z_2$ construction with OPAs (2.5) and (2.15).  We know that in this case
this group contains the original reflection
involution $r$ together with the dual
involution $i$ and other
extensions of elements of the Conway group ${\rm Co}_0$.
Furthermore, the triality symmetry \refs{\FLMb,\DGMtri}
interchanging the untwisted and twisted sectors may also still hold.
 Given this, then we can speculate
that the \auto group for $\Vp$ may be the \lq Bimonster\rq\ or
wreath square of the Monster \Atlas.
Similarly, for the other orbifold constructions, the \auto group for $\Vp$
may provide other enlargements of the Monster
which would be of obvious interest. Finally, apart from these more general
considerations,  the Monster Fricke element twisted
sectors not related to Leech lattice \autos have also to yet to be constructed
explicitly.

\beginsection{Appendix A. Modular groups in  Monstrous Moonshine.}

In this appendix we describe the modular groups relevant to the
Moonshine properties of Thompson series described by
Conway and Norton \CN.

$\GN$:   The group of matrices contained in the full modular
group of the form
$$
\eqalign{
\left(\matrix{a&b\cr
               cN&d\cr}\right),\qquad {\rm det}=1
}
\eqno(A.1)
$$
where $a,b,c,d\in Z$.

 The normaliser
\hbox {$\Nor =\{ \rho\in PSL(2,R)\vert \rho\GN\rho^{-1}=\GN\} $}, is also
required to describe Monstrous Moonshine.
Let $h$ be an integer where $h^2\vert N$  ($h^2$ divides $N$) and let $N=nh$.
Then we define the following sets of matrices.

$\Gnh$:   The group of matrices of the form
$$
\eqalign{
\left(\matrix{a&{b\over h}\cr
        cn&d\cr      }\right),\qquad {\rm det}=1
}
\eqno(A.2)
$$
where $a,b,c,d\in Z$. For $h$ the largest divisor of 24 for which $h^2\vert N$,
 $\Gnh$ forms a subgroup of $\Nor$. For $h=1$, $\Gnh=\Gn$.

$W_e$:   The set of matrices for a given integer $e$
$$
\eqalign{
\left(\matrix{ae&b\cr
        cN&de\cr}\right),\qquad {\rm det}=e \qquad e\vert\vert N
}
\eqno(A.3)
$$
where $a,b,c,d\in Z$.  $e\vert\vert N$ denotes the property that
$e\vert N$ and the greatest common
divisor $(e,N/e)=1$. The set $W_e$ forms a single coset of $\GN$ in
$\Nor$ with $W_1=\GN$.
It is straightforward to show that (up to scale factors)
$$
\eqalign{
W_e^2&=1{\rm\ mod\ }(\GN)\cr
W_{e_1}W_{e_2}&=W_{e_2}W_{e_1}=W_{e_3}{\rm\ mod\ }(\GN),\qquad
e_3={e_1e_2\over{(e_1,e_2)^2}}}
\eqno(A.4)
$$
 The coset $W_e$ is refered to as an Atkin-Lehner (AL) involution for
$\Gamma_0(N)$.  The
simplest example is the Fricke involution $W_N$ with coset representative
$\left(\matrix{0&1\cr{-N}&0\cr}\right)$ which
generates $\tau\rightarrow -1/N\tau $ and interchanges  the cusp
points at $\tau=\infty$ and  $\tau=0$.
For $e\not =n$ we can choose
the coset representative $\left (\matrix{e& b\cr N & de\cr} \right )$
where $ed-bN/e=1$ which interchanges the  cusp points at
$\tau=\infty$ and $\tau=e/N$.

$w_e$:   The set of matrices for a given integer $e$ of the form
$$
\eqalign{
\left(\matrix{ae&{b\over h}\cr
        cn&de\cr}\right),      \qquad {\rm det}=e,\qquad e\vert\vert{n\over h}
}
\eqno(A.5)
$$
 where $a,b,c,d\in Z$.  The set $w_e$ is called an Atkin-Lehner (AL)
involution for $\Gamma_0(n\vert h)$.
The properties (A.4) are similarly obeyed by $w_e$ with $\GN$ replaced by
$\Gamma_0(n\vert h)$.

$\Nor$:   The Normalizer of $\GN$ in $PSL(2,R)$ is constructed by
adjoining to $\Gnh$  all
its AL involutions $w_{e_1},w_{e_2},...$ where $h$ is the largest divisor of
24 with $h^2\vert N$ and $N=nh$.

$\Gamma_0(n\vert h)+e_1,e_2,...$ :   This denotes the
group obtained by adjoining to $\Gnh$ a particular subset
of AL involutions $w_{e_1},w_{e_2},...$ and  forms a subgroup of
$\Nor$.

\beginsection{Appendix B. Automorphism groups for twisted sectors.}
In this appendix we will derive the centraliser formula (3.19) by describing
the \auto group which preserves the OPA of $\Vorb$ where no mixing
between the various sectors $\Pa\Vb$ is considered
where  $b=a^r$ is lifted from $\b=\a^r$ of
order $m=n/r'$ with $r'=(n,r)$.  In general, $\b$ may have unit eigenvalues
(for $r'\not =1$) so that $\L$ contains a $\b$
invariant sublattice $\L_\b$ which
has  dual lattice $\L_\b^*=\L_\para\equiv {\cal P}_\b\L$.
  Likewise, we define $\L_\b^T$ to be the sublattice of $\L$
orthogonal to $\L_\b$ where the dual lattice is
$\L_\b^{T*}=\L_T\equiv (1- {\cal P}_\b)\L$.
It is then easy to show that $\L_\para/\L_\b \cong \L_T/\L_\b^T$
so that the volume of $\L_\b$ is given by
$V_\b=\vert \L_\para/\L_\b\vert ^{1/2}=
\vert \L_T/\L_\b^T\vert ^{1/2} $.

The $b$ twisted states are constructed
from a set of vertex \ops $\tilde\VL$ which form a representation of the
original untwisted OPA (2.5) with a non-meromorphic OPA.  These \ops act on
a $b$ twisted vacuum which from (3.4) we expect to have
 degeneracy $D_\b^{1/2}/V_\b$.
The construction of $\tilde\VL$ follows from
considering a string with twisted boundary condition $\tilde X(e^{2\pi i}z)=
\b^{-1} \tilde X(z)+2\pi \beta$  where
$\beta\in \L$ \refs{\Lepowsky,\KacP,\CHZn}
with a mode expansion similar to (3.7).
The corresponding states are graded by $L_0=
\sum_m \tilde \alpha_m^i\tilde \alpha_{-m}^i+{1\over 2}p_\para^2+E_0^b$ where
$p_\para$ has eigenvalues in $\L_\para$
and $E^b_0$ is the vacuum energy given in
(3.5e) which obeys $m E^b_0=0\ {\rm mod}\ 1$.
As before,  cocycle factors
$\{ c_T(\alpha)\}$ are required for a local OPA. These are defined as follows.
Consider the central extension $\hat \L$ of
$\L$ by $\langle (-1)^{m}\rho\rangle $
(where $\rho=\omega^r$, $\omega=e^{2\pi i/n}$) given by the
 following commutator \Lepowsky
$$
\eqalignno{
 c(\alpha) c(\beta) c(\alpha)^{-1} c(\beta)^{-1}=&
{\rm exp}(2\pi i S_\b(\alpha,\beta))&(B.1a)\cr
S_\b(\alpha,\beta)=-S_\b(\beta,\alpha)=
&[{1\over 2}\langle \alpha_\para,\beta_\para\rangle +
\langle \alpha_T,(1-\b)^{-1}\beta_T\rangle ]\ {\rm mod}\ 1
&(B.1b)\cr
}
$$
where $\{  c(\alpha)\}$ is a section of $\tilde \L$
and where $\alpha_\para= {\cal P}_\b\alpha\in \L_\para$,
$\alpha_T=(1- {\cal P}_\b)\alpha\in \L_T$.
(B.1b) reduces to (2.6a) when $\b=1$ and
to (3.8b) when $\b$ is without unit eigenvalues.
(B.1) also defines
a central extension $\hat\L_\b^T$ of the sublattice
 $\L_\b^T$ by $\langle \rho\rangle $
with  centre determined by the lifting of
$(1-\b)\L\subset \L_\b^T $.  Taking the
quotient of these two groups we obtain a central extension $\Lbhat$ of
$\Lb=\L_\b^T/(1-\b)\L$ by $\langle \rho\rangle $
 with centre $\langle \rho\rangle $.
$\Lb$ is a finite group of order  $\vert \L_\b^T/(1-\b)\L \vert
=\vert \L_\b^T/\L_T\vert \vert \L_T/(1-\b)\L \vert
=D_\b/V_\b^2$. In addition, $\Lbhat$ has a unique irreducible faithful
representation $\pi(\Lbhat)$ of dimension
$D_\b^{1/2}/V_\b$ in which the centre is
represented by phases $\langle \rho\rangle $ \refs{\Lepowsky,\FLMb}.
Let $T^\b$ denote the vector space on which
$\pi(\Lbhat)$ acts. Then the states $\{\vert \sigma_b\rangle \}$
of the degenerate $b$ twisted vacuum
form a basis for $T^\b$ and the cocycle factors $\{c_T(\alpha)\}$ are
$\alpha_\para$ valued matrices acting on $T^\b$ which obey (B.1).

Let us now describe the group of inequivalent \autos
${ Aut}(\tilde\VL)$ of  the OPA of
$\tilde\VL$ which act on the vector space $T^\b$.
This group is an extension of the centraliser $C(\b\vert {\rm Co}_0)$
where each lattice \auto $\g \in C(\b\vert {\rm Co}_0)$ acts on $\tilde X(z)$
in
the usual way but is lifted to a set of \autos $\{g\}$ of $\tilde \L$ where
$$
\eqalign{
g c(\alpha) g^{-1}=e^{2\pi i f_g(\alpha)}c(\g \alpha)
}
\eqno (B.2)
$$
where $f_g(\alpha)$ parameterises the liftings of $\g$.
Let $g$ and $g'$ be two inequivalent
liftings of $\g$. Then $e=g'g^{-1}$ is a lifting of the identity lattice \aut.
The group of liftings of the identity \auto form a normal subgroup of
${Aut}(\tilde\VL)$ and is parameterised by
$f_e(\alpha)$ obeying $f_e(\alpha+\beta)=f_e(\alpha)+f_e(\beta)$
and $f_e(0)=0$.
Let $\lambda^{(i)}$ be a basis for $\L$ and $\lambda^*_{(j)}$
a dual basis where
\hbox{ $\langle \lambda^{(i)},\lambda^*_{(j)}\rangle =\delta^i_j$.}
Then define $\mu^i=f_e(\lambda^{(i)})$ so that
$f_e(\alpha)=\mu^i \alpha_i=\langle \mu,\alpha\rangle $
with $\alpha=\alpha_i \lambda^{(i)}$ and
$\mu=\mu^j \lambda^*_{(j)}$ i.e. each lifting is parameterised by $\mu$.
We may determine $\mu$ by considering the inner \autos of $\hat\L$ where
$c(\beta):c(\alpha)\rightarrow {\rm exp}[2\pi i S_\b(\beta,\alpha)] c(\alpha)$
 from (B.1) and hence $\mu=-\beta_\para/2-(1-\b)^{-1}\beta_T$ for $\beta\in
\L$.
As described above, the cocycle factors $\{c_T(\alpha)\} $ used in
constructing the vertex \ops $\tilde\VL$ are defined to act on
the twisted vacuum space $T^\b$.
Hence only the inner \autos generated  by  $c_T(\L^T_\b)
=\{c_T(\beta_T)\vert\beta_T\in \L^T_\b\}\equiv \pi(\Lbhat) $,
with $\mu\in(1-\b)^{-1}\L^T_\b$, give the inequivalent liftings of the identity
to \autos of $\tilde\VL$ since $c_T(\L^T_\b)$ maps $T^\b$ onto itself.
Furthermore, from (B.2), the liftings of $\b$ itself are themselves
equivalent to liftings of the identity
lattice \auto to \autos of $\{ c_T(\alpha)\}$.
(In particular, we may define one distinguished lifting in the centre of
$\pi(\Lbhat)$,  denoted by $b={\rm exp}(-2\pi i E_0^b)\in
\langle\rho\rangle$. $b$ then describes the twisting
of the vacuum states with ${\rm exp}(2\pi i L_0)\vert \sigma_b\rangle =
b^{-1}\vert\sigma_b\rangle$).
Thus we find that the group of inequivalent
\autos ${Aut}(\tilde\VL)$  is given by
$\Lbhat.(C(\b\vert {\rm Co}_0)/\langle \b\rangle)$.

 We next describe ${Aut}(\Pa\tilde\VL)$ where
$a$ is the lifting of $\a\in C(\b\vert {\rm Co}_0)$ to an
\auto of the OPA of $\Pa\tilde\VL$ with $a^r=b$.
$a$ acts as the identity on
$\Pa\tilde\VL$ and hence each  $g\in {Aut}(\Pa\tilde\VL)$
 must commute with $a$. Therefore, $g$ is lifted from
$\g\in G_n=C(\a\vert {\rm Co_0})/\langle \a\rangle $.
The inequivalent liftings of $\g$ are
given  by the inequivalent liftings, $e$, of the identity which commute with
$a$.  Using the parameterisation above, this implies
that $\langle \mu,\alpha\rangle =
\langle \mu,\a\alpha\rangle\ {\rm mod}\ 1$ for
all $\alpha\in\L$ and hence  $\mu\in (1-\a)^{-1}\L$.
{}From above we also know that
$\mu\in(1-\b)^{-1}\L^T_\b$. Together, we find that
\hbox{$\mu\in (1-\a)^{-1}\L^T_\b$} so that
the inequivalent liftings of the identity
that commute with $a$ are given by the inner \autos generated by
$\hat K\equiv c_T(({1-\b\over 1-\a})\L^T_\b)\subseteq \pi(\Lbhat)$.
Two elements $c_T(({1-\b\over 1-\a})\alpha_T),\
c_T(({1-\b\over 1-\a})\beta_T)$ of $\hat K$ are equivalent $\iff$
$({1-\b\over 1-\a})(\alpha_T-\beta_T)= (1-\b)\lambda$ for $\lambda\in \L$
$\iff$ $\alpha_T-\beta_T=(1-\a)\lambda$ with $\lambda\in \L^T_\b$. Thus
$\hat K=m.K$,  a central extension by $\langle \rho\rangle $
of $K\equiv \L^T_\b/(1-\a)\L^T_\b$.  We therefore find that
$$
\eqalign{
Aut(\Pa\tilde\VL) =\hat K.G_n
}
\eqno(B.3)
$$
where $\hat K$ is the normal subgroup of \autos lifted
from the lattice identity \aut.

In the case where $r'=(r,n)=1$ we have $\hat K=\hat L_\a$ and so
$Aut(\Pa\tilde\VL)=\hat L_\a.G_n$.  For all the other sectors, including the
untwisted sector, the corresponding \auto group can always be expressed
as a quotient of $\hat L_\a.G_n$ by some normal subgroup.
In the untwisted case when $r=0$, the elements of $Aut(\Pa\VL)$ must commute
with $a$ and are determined  by $\mu\in (1-\a)^{-1}\L$ as above. Thus
$Aut(\Pa\VL)=L_\a.G_n=(\hat L_\a.G_n)/\langle \omega\rangle$ i.e. $Aut(\Pa\VL)$
is a quotient group of $\hat L_\a.G_n$.
For $\b=\a^r $  and $r'\not =0,1$, $\hat L_\a.G_n$  contains
a normal subgroup $\hat J=r'.J$  with  $J\equiv \L_\b/(1-\a)\L_\b$.
$\hat J$ is the group of \autos of $\Pa\Va$ lifted from the identity
lattice \auto and given by the inner \autos generated by $c_T(\L_\b)$
(Note that the commutator (B.1) in this sector is determined by
$S_\a(\alpha,\beta)=\langle \alpha,(1-\a)^{-1}\beta\rangle$).
We therefore find that  $Aut(\Pa\tilde\VL)=\hat K.G_n=(\hat L_\a.G_n)/ \hat J$.
Thus for all sectors $\Pa\tilde\VL$, including the untwisted one,
we may describe the OPA \auto group $Aut(\Pa\tilde\VL)$
by $\hat L_\a.G_n$ where some normal subgroup may act as
the identity on $\Pa\tilde\VL$, namely $\langle \omega\rangle$ for $r=0$ and
$\hat J$ for $r'\not=0, 1$.

The \ops of $\Pa\tilde\VL$ create $b$ twisted states from the twisted
vacuum vector space $T^\b$. We may then define vertex \ops $\{\psi_b\}=
\Pa\Vb$ which create these states from the untwisted vacuum where
(schematically) $\tilde\phi \sigma_b\sim \psi_b$
with $\tilde\phi\in \Pa\tilde\VL$
and $\sigma_b$ creates a twisted  vacuum state.
This OPA algebra is also invariant under $\hat L_\a.G_n$ with an appropriate
identity action under a normal subgroup as described above.
Likewise, the intertwining
OPA  between the various twisted sectors $\Pa\Vb$ as in (3.14),
which is expected to exist,  is invariant under $\hat L_\a.G_n$.
Note that we are not considering here mixing
(triality) \autos  between the various sectors which are expected
as in the usual Moonshine constructions \refs{\FLMb,\DGMtri,\DongMason}.
We therefore find that the OPA of
$\Vorb=\Pa(\VL\oplus\Va\oplus ...{\cal V}_{a^{n-1}})$
is invariant under  $\hat L_\a.G_n$
where no mixing between the various twisted sectors is considered. With
$a^*$ defined on $\Vorb$ as in \S 3 (the \ops of $\Vak$ are eigenvectors with
eigenvalue $\omega^k$) we have
$$
\eqalign{
C(a^*\vert \Morb)=\Lahat.G_n
}
\eqno (B.4)
$$
where $\Morb=Aut(\Vorb)$ is the complete \auto group for $\Vorb$. This is
the result given in (3.19) for the 38 modular invariant orbifold constructions
from the lattice \autos of Table 1.

We may also compute the centraliser $C(g_n\vert \Morbh)$
where $g_n$ is lifted from one of the 13  lattice
\autos $\a$ of Table 2.  Orbifolding $\VL$ with respect  to
$\a'=\a^h$ gives a modular consistent
construction $\Vorbh$ and $g_n^h=a'^*$ where $a'^*$
is dual to the lifting of $\a'$.  From  (B.4)
we have $C(g_n\vert \Morbh)\subset C(a'^*\vert \Morbh)=\hat L_{a'}.G_{n'}$
where $G_{n'}=C(\a'\vert{\rm Co}_0)/\langle \a'\rangle$.
We may next repeat most of the argument given above to firstly
find the \auto group for the OPA of the vertex \ops
$\Pap\Vap$.  Each \auto $g\in C(g_n\vert Aut(\Pap\Vap))$
is lifted from a lattice \auto $\g\in C(\a\vert{\rm Co}_0)=n.G_n$
 where the inequivalent liftings are determined by
the group of liftings of the identity lattice \auto
which commute with  $g_n$.
This forms a normal subgroup of $Aut(\Pap\Vap)$, as
before,  generated by the inner \autos with respect to
$c_T(({1-\a^h\over 1-\a})\L)\subset \pi(\hat L_{a'})$. This group
together with $g_n$ itself generates $\Lahat$.
Thus the group of \autos of $\Pap\Vap$  that commute with
$g_n$ is $\Lahat.G_n$.
By following an argument similar to that above, we can also show that
 the \autos of $\Pap\Vbp$ which commute with $g_n$
are given by the quotient group of $\Lahat.G_n$ by a normal subgroup.  Thus
the centraliser is $C(g_n\vert \Morbh)=\Lahat.G_n$ as in (3.19).

\vfill\eject
\vbox{\tabskip=0pt \offinterlineskip
\def\tabrule{\noalign{\hrule}}
\halign{
% \hfil$#$\hfil \  &
% \hfil$#$\hfil\  & \hfil$#$\hfil\  &
% \hfil$#$\hfil\  & \hfil$#$\hfil\  &
% \hfil$#$\hfil\  & $#$\hfil\
 \vrule \  \hfil$#$\hfil \  & \vrule \  \hfil$#$\hfil \  &
\vrule \  \hfil$#$\hfil \  & \vrule \  \hfil$#$\hfil \  &
\vrule \  \hfil$#$\hfil \  & \vrule \  \hfil$#$\hfil \  \vrule
\cr\tabrule
 \a\in{\rm Co_0}   &\Gamma_a  & \hat L_\a & G_n & C(g_n\vert M) & \cr
\tabrule
{2^{24}/1^{24}}  &  2-  & 2^{1+24} & \Co  & 2^{1+24}.\Co& \dagger\cr
{3^{12}/1^{12}}  & 3-  & 3^{1+12} & 2.{\rm Sz} &
3^{1+12}.2.{\rm Sz}& *\cr
{4^8/1^8}  &    4-  & 4.4^{8} & 2.2^6.S_6(2) &
4.2^{15}.2^8.S_6(2)& \ddagger  \cr
{5^6/1^6}& 5-  & 5^{1+6} & 2.{\rm HJ} & 5^{1+6}.2.{\rm HJ}  &  * \cr
{2^6 6^6/1^6 3^6}  & 6+3 &  2^{1+12}\times 3 & 3.U_4(3).2 &
2^{1+12}.3^2.U_4(3).2&  \dagger\cr
{3^4 6^4/1^4 2^4} &   6+2 &  2\times 3^{1+8} & 2^{1+6}.U_4(2) &
2.3^{1+8}.2^{1+6}.U_4(2)& * \cr
{2.6^5/1^5 3} &  6-  &  2^{1+6}\times 3^{1+4} & 2.U_4(2) &
2.3^{1+4}.2^{1+6}.U_4(2)&  \ddagger \cr
{7^4/1^4}  &  7- &  7^{1+4} & 2.A_7 & 7^{1+4}.2.A_7& *\cr
{2^2 8^4/1^4 4^2}  &   8- &  8.(8^2\times 4^2) & [2^9. 3] &
[2^{22}. 3] & \ddagger\cr
{9^3/1^3}  & 9-  & 9.(9^2\times 3^2) & [2^4. 3^3] & [2^4. 3^{11}]& *\cr
{2^4 10^4/1^4 5^4}  &  10+5 &  5\times 2^{1+8} & (A_5\times A_5).2 &
5\times 2^{1+8}.(A_5\times A_5).2&  \dagger\cr
{5^2 10^2/1^2 2^2}  &10+2 &  2\times 5^{1+4} & 2^{1+4}.A_5 &
2.5^{1+4}.2^{1+4}.A_5&  *\cr
{2.10^3/1^35} &  10- &  2^{1+4}\times 5^{1+2} & 2A_5 &
2.5^{1+2}.2^{1+4}.A_5&  \ddagger \cr
{2^4 3^4 12^4/1^4 4^4 6^4}  &  12+4 &
4\times 3^{1+4} & 2.2^4.S_6 & [2^{11}.3^7. 5]&  \ddagger\cr
{4^2 12^2/1^2 3^2} &  12+3 &  4.4^4\times 3 & [2^5. 3^2] &
[2^{15}. 3^3]&  \ddagger \cr
{2^2 3.12^3/1^3 4.6^2}  &  12-  & 4.4^2\times 3^{1+2} & [2^4. 3] &
[2^{10}. 3^4]&  \ddagger\cr
{13^2/1^2} & 13- &  13^{1+2} & 2.A_4 & 13^{1+2}.2.A_4 & *\cr
{2^3 14^3/1^3 7^3} &  14+7 &  7\times 2^{1+6} & L_2(7) &
[2^{10}. 3. 7^2]&  \dagger \cr
{3^2 15^2/1^2 5^2} & 15+5 &  5\times 3^{1+4} & 2.A_5
& [2^3.3^6.5^2] & *\cr
{2.16^2/1^2 8}  &  16- &  16.8^2 & [2^3] & [2^{13}]&  \ddagger\cr
{9.18/1.2} & 18+2  & 2\times 9.9^2 & [2^3 .3] & [2^4 .3^7]& * \cr
{2^3 3^2 18^3/1^3 6^2 9^3} &  18+9 & 2^{1+4}\times 9 &
[2.3^3] & [2^6.3^5]&  \dagger\cr
{2.3.18^2/1^26.9}  & 18- &  2^{1+2}\times 9.3^2 & 2.3 &
[2^4. 3^5]&  \ddagger\cr
{2^2 5^2 20^2/1^2 4^2 10^2}  &  20+4 &  4\times 5^{1+2} &
2.S_4 & [2^6.3.5^3]&  \ddagger\cr
{7.21/1.3} & 21+3 &  3\times 7^{1+2} & 2.3 & [2.3^2.7^3] & *\cr
{2^22 2^2/1^2 11^2} &  22+11 &  2^{1+4}\times 11 & S_3 &
[2^6.3.11]&  \dagger \cr
{2.3^2 4.24^2/1^2 6.8^2 12} &  24+8 & 8\times 3^{1+2} & [2^4] &
[2^7.3^3]&  \ddagger \cr
{4.28/1.7} &  28+7 &  4.4^2\times 7 & 2 & [2^7.7] &  \ddagger\cr
{2^3 3^3 5^3 30^3/1^3 6^3 10^3 15^3} &  30+6,10,15 &
2\times 3 \times 5 & A_6 & [2^4.3^3.5^2] &  \dagger\cr
{2.6.10.30/1.3.5.15} & 30+3,5,15 &  2^{1+4}\times 3\times 5
& S_3 & [2^6.3^2.5]&   \dagger \cr
{2^23.5.30^2/1^2 6.10.15^2} &  30+15 &
2^{1+2}\times 3\times 5 & 2 & [2^4.3.5]&  \dagger \cr
{3.33/1.11} & 33+11 &  3^{1+2}\times 11 & 2 & [2.3^3.11]& * \cr
{2.9.36/1.4.18} &  36+4 &  4\times 9.3^2 & 2 & [2^3.3^4] &  \ddagger\cr
{2^2 3^2 7^2 42^2/1^2 6^2 14^2 21^2} & 42+6,14,21 & 2\times 3 \times 7 &
A_4 &  [2^3.3^2.7] &   \dagger\cr
{2.46/1.23} & 46+23 &  2^{1+2}\times 23 & 1 & [2^3.23] &   \dagger\cr
{3.4.5.60/1.12.15.20}& 60+12,15,20 &  4\times 3
\times 5 & 2 & [2^3.3.5] &   \ddagger \cr
{2.5.7.70/1.10.14.35}  & 70+10,14,35 &  2\times 5 \times 7 & 1 &
[2.5.7]&   \dagger\cr
{2.3.13.78/1.6.26.39} & 78+6,26,39 &  2\times 3\times 13 & 1 &
[2.3.13] &    \dagger\cr
\tabrule} }

\bigskip
\centerline{\bf Table 1}
\medskip
The 38 conjugacy classes of $\rm Co_0$ obeying (3.6). The first column gives
$\g$ in Frame shape notation. The corresponding modular group $\Gamma_a$
appears in column 2. The groups appearing
in columns 3, 4 and 5 are expressed in
terms of standard Atlas groups \Atlas\ where $n^k$
denotes the direct product of $k$ cyclic groups of order $n$ and
$[p_1^a.p_2^b...]$ denotes an unknown group of the given order.
$A\times B$ denotes a direct product group and $A.B$ denotes a group with
normal
subgroup $A$ where $B=A.B/A$.

\vfill\break
\vbox{\tabskip=0pt \offinterlineskip
\def\tabrule{\noalign{\hrule}}
\halign{
% \hfil$#$\hfil \  &
% \hfil$#$\hfil\  & \hfil$#$\hfil\  &
% \hfil$#$\hfil\  & \hfil$#$\hfil\  &
% \hfil$#$\hfil\  & $#$\hfil\
\vrule \  \hfil$#$\hfil \  & \vrule \  \hfil$#$\hfil \  &
\vrule \  \hfil$#$\hfil \  & \vrule \  \hfil$#$\hfil \  &
\vrule \  \hfil$#$\hfil \  \vrule
\cr\tabrule
\a\in{\rm Co_0} &  \Gamma_a  &  \hat L_\a & G_n & C(g_n\vert M)\cr
\tabrule
{4^{12}/2^{12}} &  4\vert 2- & 4.2^{12} & G_2(4).2 &
4.2^{12}.G_2(4).2\cr
{6^8 /3^8} &  6\vert 3- &  3\times 2^{1+8} & A_9 &
3\times 2^{1+8}.A_9\cr
{8^4/2^4} &  8\vert 2- &  8.4^4 &  2.2^4.A_6 & 8.2^9.2^4.A_6\cr
{8^6/4^6} &  8\vert 4- &  8.2^6 & U_3(3) & 8.2^6.U_3(3) \cr
{6^2 12^2 /2^2 4^2} &   12\vert 2+2 &   4\times 3^{1+4} &
[2^7. 3] & [2^9. 3^6]\cr
{12^4/6^4} &  12\vert 6- &  3\times 4.2^4 & A_5\times 2 &
[2^9.3^2 5]\cr
{15^2/3^2} &   15\vert 3- &  3\times 5^{1+2}
& 2.A_4 & [2^3.3^2.5^3]\cr
{4^2 20^2/2^2 10^2} &  20\vert 2+5 &  4.2^4\times 5 & A_5 &
[2^8.3.5^2]\cr
{8.24/2.6} &  24\vert 2+3 &  3\times 8.4^2 & [2.3] &
[2^8.3^2]\cr
{12.24/4.8} &  24\vert 4+2 &  8\times 3^{1+2} & [2^2] &
[2^5.3^3]\cr
{24^2/12^2} &  24\vert 12- &  3\times 8.2^2 & 3 & [2^5.3^2]\cr
{6.42/3.21} &  42\vert 3+7 &  2^{1+2}\times 3\times7 & 1 &
[2^3.3.7]\cr
{4.6.14.84/2.12.28.42} &  84\vert 2+6,14,21 &
4\times 3\times 7 & 1 & [2^2.3.7]\cr
\tabrule}}

\bigskip
\centerline{\bf Table 2}
\medskip
 The 13 conjugacy classes of $\rm Co_0$ obeying (3.6a) only.
 For such each $\a$ there is an integer $h\vert n$, $h\vert 24$ where
 $\a^h$ appears in Table 1.

\listrefs
\end